\pdfoutput=1
\documentclass[prl,twocolumn,showpacs,preprintnumbers,amsmath,amssymb,floatfix,reprint]{revtex4}
\usepackage{hyperref}
\usepackage{epsfig}
\usepackage{color}
\begin{document}

\title{Electrical charging effects on the sliding friction of a model nano-confined ionic liquid}

\author{R. Capozza$^{1,2}$, A. Benassi$^{2,3}$, A. Vanossi$^{2,1}$, and E. Tosatti$^{1,2,4}$}
\affiliation{
$^1$ International School for Advanced Studies (SISSA), Via Bonomea 265, 34136 Trieste, Italy \\
$^2$ CNR-IOM Democritos National Simulation Center, Via Bonomea 265, 34136 Trieste, Italy \\
$^3$ Institute for Materials Science and Max Bergmann Center of Biomaterials, TU Dresden, 01062 Dresden, Germany\\
$^4$ International Centre for Theoretical Physics (ICTP), Strada Costiera 11, 34014 Trieste, Italy}

\begin{abstract}
Recent measurements suggest the possibility to exploit ionic liquids (ILs) as smart lubricants 
for nano-contacts, tuning their tribological and rheological properties by charging the sliding interfaces. 
Following our earlier theoretical study of charging effects on nanoscale confinement and squeezout of a 
model IL, we present here molecular dynamics simulations of  the frictional and lubrication properties 
of that model under charging conditions.
First we describe the case when two equally charged plates slide while being held together to a confinement 
distance of a few molecular layers. The shear sliding stress is found to rise strongly and 
discontinuously as the number of IL layers decreases stepwise. However the shear stress shows, 
within each given number of layers,  only a weak dependence upon the precise value of the normal load, 
a result in agreement with data extracted from recent experiments.
We subsequently describe the case of opposite charging of the sliding plates, and follow the shear stress 
when the charging is slowly and adiabatically reversed in the course of time, under fixed load. 
Despite the fixed load, the number and structure of the confined IL layers  changes with changing charge, 
and that in turn drives strong friction variations. The latter involve first of all charging-induced freezing of 
the IL film,  followed by a discharging-induced melting, both made possible by the nanoscale confinement. 
Another mechanism for charging-induced frictional changes is a shift of the plane of maximum shear from 
mid-film to the plate-film interface, and viceversa. While these occurrences and results invariably depend 
upon the parameters of the model IL  and upon its specific interaction with the plates, the present study 
helps identifying a variety of possible behavior, obtained under very simple assumptions, while connecting 
it to an underlying equilibrium thermodynamics picture.
\end{abstract}
\date{August 18, 2015}
\maketitle


\section{Introduction}
Ionic liquids (ILs) -- organic salts that are liquid at room temperature -- are 
of considerable and increasing physical and technological interest.
ILs are characterized by negligible vapor pressures, high temperature stability and a wide 
electrochemical window \cite{plechkova08}. Moreover their physical properties can 
be widely tuned by changing the molecular structure of the cation-anion pairs \cite{hallett11}.
Many ILs strongly adhere to solid surfaces \cite{hayes10} and can form wear-protective 
films capable of resisting much higher loads than those of molecular lubricants \cite{perkin10}. 
For this reason among others their use as boundary lubricants has been recently pursued in experimental 
studies under nanoscale confinement, employing both the surface force apparatus (SFA) \cite{perkin10,spencer},
and atomic force microscopy (AFM) \cite{werzer12,li13,bennewitz,li14,hoth14}, as well in
in a number of computer simulations at various levels of idealization \cite{padua13, federici14,fajardo15}.
Our specific interest focuses here on the exploration of the frictional changes induced by electrical 
charging of the plates which confine the IL, a subject only partially covered by previous theoretical 
works ~\cite{kornyshev, fajardo15, capozza2015}. 
Besides lubrication, the dynamical behavior of ILs under charging is of additional interest in the field 
of supercapacitors ~\cite{kim11, shim10}. 
Experimentally, there is a number of nano-frictional studies of 
surfaces charged with respect to a reference electrode~\cite{bennewitz, black13, li14}. Not so  
between two oppositely charged electrodes. Recent macroscopic friction studies have begun to appear for
IL sliding under charging conditions ~\cite{dold2015,yang14}. Here however we shall restrict ourselves to
friction under conditions of nanoscale confinement.\\
At the molecular level, ILs confined between hard plates become structured in the form of layers that run 
parallel to the plates, not unlike other liquids but further characterized by the charge order typical 
of molten salts, with an alternation of positive and negative ion layers, and an interlayer separation 
that corresponds to the ion pair size \cite{perkin10,perkin,werzer12,hoth14,atkin07,bou-malham10,ueno10,
perkin11,perkin12,li13b}. Plate charging is expected to give rise to rearrangements of the IL layering 
and ordering, at least near the confining plate-IL interfaces, with a consequent and probable change of the
lubrication properties under plate sliding. Charging-induced friction changes have already  been to some extent detected 
and described experimentally \cite{bennewitz,li14}. Not unexpectedly, they are reported to depend 
on the particular kind of IL and of confining surfaces chosen.  
For example, a strong reduction of the friction of an AFM silica colloid probe was found by negatively biasing a 
Au(111) surface \cite{bennewitz} immersed in $[Py_{1,4} FAP]$. However, a similarly strong friction drop 
occurred at positive bias upon sliding a sharp AFM tip on highly oriented pyrolytic graphite immersed in a 
$[HMIm] FAP$ IL \cite{li14}.

This diversity of behavior, reflecting the variety of ILs and the different nature of the confining and sliding 
surfaces, suggests some flexibility in the corresponding theoretical modeling, aimed at a broad exploration of how some 
of this variety of behavior could be addressed with a handful of model parameters, rather than describing in details one or another particular case. 
The minimal IL model and a natural first choice is a simple molten salt, such as liquid NaCl. Fedorov and Kornyshev ~\cite{kornyshev} 
did that, their model consisting of two spherical Lennard-Jones (LJ) particles with unequal radii further endowed 
with opposite charges. The frictional behavior of that model was recently investigated by Fajardo et al.~\cite{fajardo15} 
who noted interesting similarities with data by Li et al.~\cite{li14}. However, a parallel study which we conducted of 
this type of model under simple confinement and squeezout ~\cite{capozza2015} revealed that the 
confined liquid film evolved far too sharply from a layered liquid for neutral plates at large spacings, to a strongly 
crystallized rocksalt structure solid under narrow confinement and/or plate charging. 
On top of that, the simple molten salt model wetted the plates much too readily, even given a reasonable choice of parameters. 
Both of these aspects, excessive crystallization and excessive wetting tendencies, represent a severe 
oversimplification by comparison with real IL lubricants whose wetting capability is generally more modest, and where, 
owing to far more complex molecular structures, the layering and solidification which occurs under strong confinement 
does not usually imply cystallization, and is replaced by glass-like disordered structures. 
That complex behavior is of course much better captured by fully realistic IL model simulations, of which there are
several good examples in literature,~\cite{padua13, federici14} but whose difficulty and intrinsic complexity make the studies much 
more work-intensive while at the same time emphasizing aspects that are to some extent undesirably specific to each case. 

A previous study of our own IL model~\cite{capozza2015} was designed to bridge the gap between these two extremes, either oversimplification 
or overspecificity. We showed that a minimal modification of the charged LJ  model, consisting of a  neutral "tail" rigidly 
attached to one of the ions (the cation in our case, but the anion could have been equivalently chosen) could go a long 
way in our desired direction, making the structural and squeezout properties 
of the confined IL model considerably more realistic while still not specific. While the anion and the charged part of the
cation retain a strong Coulomb correlation, the neutral tail is generally and merely pushed out of the way, resulting
in poor tail-tail correlations. The tail-induced disorder, along with the steric hindrance they introduce, impede 
overall rocksalt crystallization with drastic changes in the ILs solidification, also introducing new unexpected 
structures as well as the desired glassy aspects and wettability features. These structural features were found to evolve and change under plate 
charging, foreshadowing a variability that could be explored through a variation of the model parameters. At the same time, this
IL "tailed model" (TM), to be further detailed below, is simple enough to permit a novel, quantitative calculation 
and characterization of the confined IL enthalpy versus interplate distance, and -- of crucial importance here -- its variations 
under plate charging. The minima of the enthalpy-distance curves identify stable or metastable layering states of the confined IL, 
showing a perfect fit with the evolution in the structural layering oscillations of density and charge observed
in simulations of the confined IL under variable interplate separation and plate charge. 
The layering transitions during squeezout, well established in real IL experiments ~\cite{atkin07, smith13, spencer}, are  
reproduced by the TM  model and directly related to jumps between enthalpy minima, while these minima evolve 
and cross values under closing of the gap between plates, or under variable plate charging ~\cite{capozza2015}. 
With that very useful characterization, the TM model provides a ready tool for a fresh study of some generic effects of 
electrical plate charging on the nanoscale sliding friction of confined ILs.\\

In this work we first introduce and detail the TM model and the chosen confining geometry. We then analyze the sliding simulations
and discuss how the structural changes, induced by a negative plate charging, modify the frictional properties. 
Interesting observations emerge examining the shear velocity profile inside the IL film as a function of the plate charging.
First, we observe a switching of the shear band plane from the plate-film interface, typical of low 
friction, to the mid-film region, typical of larger friction. A second important effect is the
charging-induced solidification/melting process described in our previous, static study, and typically 
ignored in literature. Charged plates generally increase the IL wetting, giving rise to  "electrowetting". 
But, as most ILs are not far above freezing at room temperature,
increased wetting will in turn encourage solidification, the capability to withstand load, with a large effect on friction.
Conversely, plate neutrality and partial dewetting facilitates melting of the IL film, leading to squeezout under even
moderate load.\\ 

The final and most specific part of this work is the investigation of the effects on friction 
of opposite charging (equivalent to an applied interplate voltage) of the two confining plates. A given charge state of the plates
is accompanied by a strong structuring of the IL near and between the plates. A reversal of plate charges is followed by an accompanying
destructuring and even melting of that layering structure, prior to the reversal of local IL ordering and eventual re-solidification. The  
sliding friction between plates undergoes a corresponding charge-dependent evolution as a consequence. This kind of phenomena,
only poorly explored so far, constitute a fresh prediction of this study.

\section{Model, simulation details, system preparation}
%
\begin{figure}
\centering
\includegraphics[width=8.5cm,angle=0]{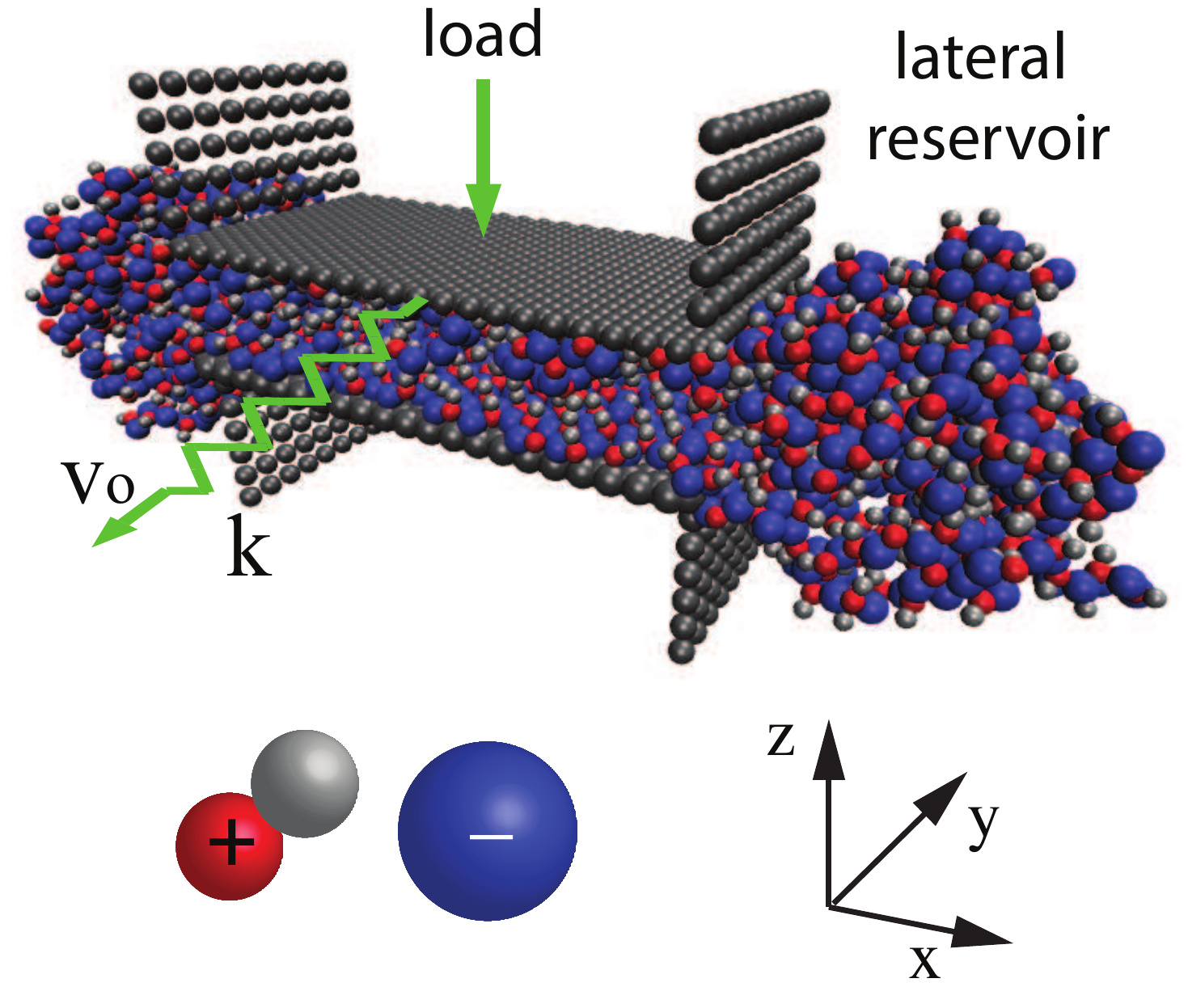}
\caption{(Color online) Sketch of the simulation geometry with open boundaries along $x$ and $z$ directions and 
periodic boundaries along $y$, i.e. the direction along which the sliding force is applied through a spring moving 
with speed $V$, a situation conceptually similar to an SFA setup. The width $L$ of the plates along $x$ is $20$ nm. 
By applying an external $z$-oriented force to the top plate, both at $V=$0 and at $V>$0, the size $D$ of the gap between 
the plates can change since the liquid can flow in and out the lateral reservoirs. The shapes of the anion 
(blue) and cation (red) are sketched in the lower left corner.}
\label{fig1}
\end{figure}
%
To pursue our desired broad scenario of possible charging effects on the lubrication properties of nano-confined ILs we 
adopt the simple TM model ~\cite{capozza2015} which deliberately neglects the fine structure of the ions retaining 
only minimal features common to most of the ILs, 
such as the anion-cation asymmetry, their possible amphiphilic character and their generally irregular shape.
The TM model is a two-component charged  Lennard-Jones (LJ) system where anions and cations 
not only have different radii but where a tail is attached to the cation. The cation is therefore a dimer consisting of 
a positively charged, small-sized LJ head, rigidly bound to an equal size, neutral particle, see Fig. \ref{fig1}.
The effect of the tail upon the ILs wetting and confinement properties is important, as mentioned in the 
introduction and recently shown in ref. \cite{capozza2015}, where the details about the adopted 
potentials and the corresponding parameters are also specified and discussed. Essentially the tails give the 
molecules a larger gyration ratio in the liquid state, and
prevent low temperature ordering replacing crystallization with a glass-like solidification -- all  features that are 
common and important in real ILs.
All the molecular dynamics (MD) simulations were performed using the LAMMPS code~\cite{lammps}. 
The long range Coulomb interactions have been treated in reciprocal space using a particle-particle particle-mesh solver.
Canonical ensemble configurations were generated by means of a Langevin thermostat 
directly applied to the lubricant molecules. The plates were treated as rigid bodies, 
the lower one fixed and the upper one subjected to a $z-$directed load force $F_N$ and driven to slide along (x,y).
Their structure consists of rigid close-packed triangular lattices of LJ 
particles with nearest neighbor spacing 0.52 nm, mimicking the periodicity of a mica surface. A graphite, or graphene
plate, which unlike mica may be electrically charged ~\cite{li14} has a
periodicity of about 0.246 nm, which is not too different from half that value.
In most of our simulations the liquid is confined between 
two identical plates with a modest surface charge density 
$q=-4\mu C/cm^2$, subsequently increased to $q=-12\mu C/cm^2$.
As a reference, the surface charge density quoted for mica in contact with H$_2$O is -33.4 $\mu C/cm^2$, 
a much larger value ~\cite{northern91}. 
The state of charge of a mica surface in contact with an IL will certainly differ, 
but it is still generally believed that some charging remains.
Since nature as well as the algorithm used to treat the long range interaction in reciprocal space require the overall 
system to be neutral, when endowing both plates with a total net nonzero charge we correspondingly
add extra ions of the opposite charge to compensate.\\
The lateral drive is actuated through a spring $k$ connected to the top plate and pulled at constant velocity $V$. 
The same value $V=2.2m/s$  was adopted throughout unless another value is explicitly stated. 
The instantaneous frictional force opposing the motion is measured through the 
elongation of the driving spring as $F_L(t)=k(V t-Y_{CM}(t))$,  where $Y_{CM}(t)$ is the $y$ 
coordinate of the top plate center of mass. The relevant quantity in friction is eventually the shear 
stress $\sigma = <F_L>/A$ where $A$ is the area of the contact. In our case $A = 177$ nm$^2$ is a constant, so  
presenting $<F_L>$ or $\sigma$ is just the same. In experiments however the contact area is not generally
constant, so that the shear stress must be extracted before making comparisons.\\  
The open geometry described in  Fig. \ref{fig1} has been chosen because it permits 
particles to be squeezed out or sucked in from two lateral IL droplets. These droplets  thus
serve as liquid reservoirs allowing the number of ions effectively confined inside the gap to adjust and change  
dynamically depending on  the loading, charging, and sliding conditions, realizing a kind of effectively grand-canonical configuration.
The distinguishing feature of the open boundary geometry employed here, as opposed to several previous studies
which used closed periodic boundary conditions, is to address a situation much closer to real SFA or AFM experiments, 
permitting in particular squeezeout transitions and the consequent transverse ($z$) and in-plane $(x,y)$ reordering.\\

Before the sliding simulations begin, we prepared equilibrium starting configurations. In our protocol, the starting configurations 
were obtained by filling the open gap between  initially distant plates, in any desired state of charge,  with
IL, and then reducing the gap width $D$ with $\dot{D}$ = 0.11 m/s, generally slow enough
to enact a reasonably adiabatic squeezout, as was found in Ref.\cite{capozza2015}. 
The adiabaticity is facilitated by 
our chosen temperature  $T= 225K$, substantially higher the TM bulk melting temperature $T_{melt}\simeq 150K$~\cite{capozza2015},
but as will be discussed below it is not automatically guaranteed, because both confinement and charging favor an increase
of viscosity and a tendency to solidify.  Above a minimal interplate distance $D_c\simeq 4nm$ and for small plate charge, the simulated 
IL is completely liquid, even if structured in
close vicinity of the plates owing to their rigid wall nature and also to their charging. Interestingly, this near-plate local structure 
develops not only along $z$, i.e., normal to the interface, but also along $(x,y)$, {\it parallel} to the interface. Parallel IL ordering, rather more difficult to
detect experimentally, is currently being pursued and detected with AFM tools~\cite{segura13, elbourne15}. As $D$ is
further and gradually decreased, the confinement effects increase, the liquid-like diffusion diminishes and below a critical film thickness $D_c$ there is an 
effective interlocking of the structure emanating from the two facing plates, leading to a freezing of the IL into a solid-like layered 
arrangement with an odd number $N_{layer}$ of alternating charge layers -- when the two plates are neutral or identically charged. 
While no claim can obviously be made to represent a realistic situation, we note that IL confinement between  
mica plates, typically used in SFA friction experiments, is known to behave precisely like that ~\cite{perkin10, atkin07}. 

\begin{figure}
\centering
\includegraphics[width=8.5cm,angle=0]{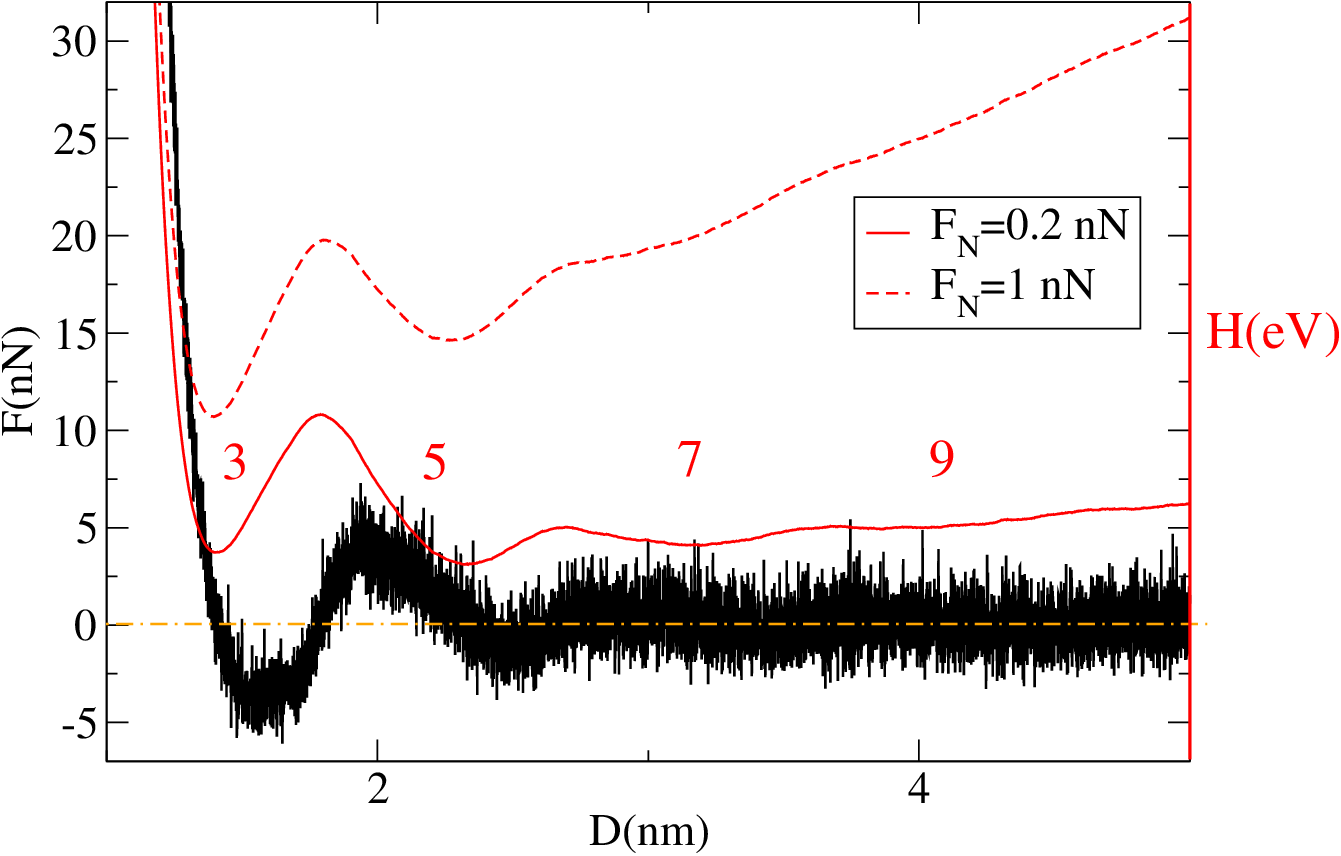}
\caption{(Color online) Interplate force $F(z)$ as a function of distance, 
and free enthalpy $H(D)$ calculated by Eq.(1) in the absence of sliding and 
for indicated values of normal load $F_N$ and an equal charge on both plates 
of $q=-4 \mu C/cm^2$.}
\label{fig2}
\end{figure}

As was shown in the previous paper~\cite{capozza2015} our choice of scheme and geometry 
permits additional thermodynamical insights. Treating the interplate separation 
$D$ as the external control variable, a free enthalpy $H(D,F_N,q)$
can be calculated first estimating the free energy $W(D)$ by integration of the average force $<F(D)>$ 
measured between the plates as their distance $D$ is reduced. 
\begin{equation}
W(D)  = \int_D^\infty <F(z)> dz
\end{equation}
where $<F(z)>$ is the average z-oriented force exerted by the IL on the plates while they are a 
distance $z$ apart. After that the free enthalpy at given load and charge can be obtained as
\begin{equation}
H(D,F_N,q)=W(D) + F_N D,
\end{equation}
Free enthalpy curves are displayed in Fig.\ref{fig2} for two values of normal load $F_N$,
and predict the load-induced transition $N_{layer}\rightarrow N_{layer}-2$ layers. 
The plate-plate free enthalpy generally displays
as a function of $D$ several odd-$N$ minima whose relative values depend on plate charging and load. 
Once specified charge and load, one of 
them is the absolute minimum, therefore the thermodynamically stable layered state, and all others are higher,
metastable states, which may still be very long lived. If the layered state is obtained by applying a force rather than by
forcing a distance $D$, the precise number of layers attained in the course of a given simulation (or experiment) will be
history-dependent, in that more than one observed thickness may correspond to the same applied force.  
This picture is quite close to what is observed in SFA and AFM  ~\cite{perkin10, atkin2, spencer}.  
In addition to the alternating charge layer structure, we had previously found that the 
confinement-induced solid structure of the TM model, although far from completely cystalline, 
carried a second partly crystalline feature, consisting of vertical, neutral planes, inside 
which the anions and cations arrange in a roughly square lattice, albeit with totally 
disordered tail orientations~\cite{capozza2015}. Further decrease of $D$ caused this nearly solid IL 
to squeeze out by successive pairs of ion planar layers, positive and negative, thus preserving local charge neutrality, 
reducing the extension of vertical planes.\\ 
Starting with large $D$ and an initially fluid IL between the plates, the force $<F(D)>$ resisting squeezout, 
initially zero, rises as $D$ drops. 
The IL is partly squeezed out, and the part that remains 
trapped acquires a structure consisting of $N_{layer}$ layers (odd for equally 
charged plates, even for opposite charging). 
The squeezout transitions $N_{layer} \rightarrow N_{layer}-2$ are clear first order transitions
in this picture, the free enthalpy jumping across barriers between successive $H(D)$ local minima.
Decreasing $D$, and thus increasing $<F(D)>$ the lowest enthalpy state moves to a 
lower and lower number of layers. Each of these states with progressively lower, integer number 
of IL layers provides a starting state for our successive sliding simulations. 

Upon checking, we actually found that configurations generated by fast squeezouts often possess internal layers 
that are only partly populated. This kind of ill-prepared states
can survive as metastable for a long time. 
When faced with these ill-prepared states, initial particle configurations were further refined
until a sufficiently reliable equilibrium particle population was reached. That allowed us to 
discard as a rule metastable initial configurations, and to describe the sliding behaviour 
of the stable, history independent ones, striving to discard metastable states as much as possible 
within our modest simulation durations.  
%
\section{Simulated sliding}

As mentioned earlier, there are two types of plate charging relevant for sliding friction. The first type, and the
one generally realized experimentally, is single-plate charging relative to an electrode in electrical contact with the IL.
In this case the aspect that matters is the IL structuring close to the charged plate, whereas the second plate, or tip, 
often insulating, acts as a merely mechanical probe of that structure. We replace this situation by simulating two equally charged plates,
neutralized by adding the right amount of extra ions to the IL.  The second type, whose experimental feasibility we ignore, 
but which looks even more interesting, is that of opposite plate charging. This  we simulate without difficulty,
restricting the charge magnitude to values small enough to correspond to an interplate voltage below ~5 V.

We  start off our sliding simulations with equally charged plates, endowed with a negative charge density $q=-12 \mu C/cm^2$ 
(about 1/3 of the nominal value quoted for mica in experimental conditions) and  with the  IL confined between them in a variety of 
$N_{layer}$ configurations, obtained as a result of squeezout down to different $D$ values, as detailed in the previous section, 
further equilibrated under an applied load $F_N$, chosen to be close to $<F(D)>$.   Sliding friction depends upon the 
IL layer number $N_{layer}$ and the  load $F_N$. 
  
\begin{figure}
\centering
\includegraphics[width=8.5cm,angle=0]{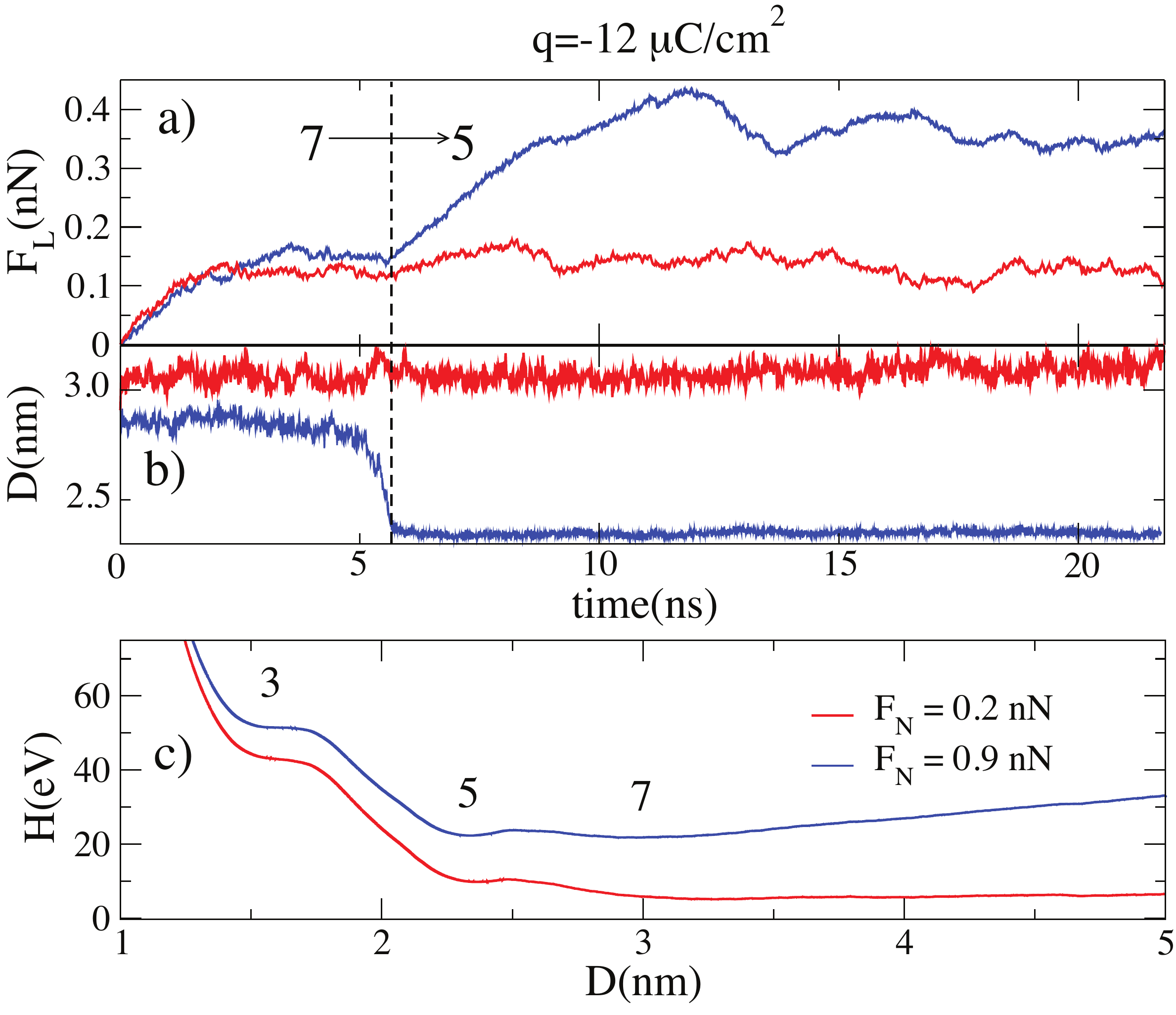}
\caption{Time evolution of a) the frictional force $<F_L>$ and b) interplate gap width $D$ as a 
function of time for two normal loads $F_N=0.2$ nN (stable) and $ F_N =0.9$ nN (metastable). 
c) Free enthalpy curves from squeezout before sliding for indicated values of $F_N$. 
Note a stronger local minimum at $N_{layer} =5$ for $ F_N =0.9$ nN, justifying the  
$7 \to 5$ transition under sliding.}
\label{fig3}
\end{figure}

Fig.\ref{fig3}a illustrates two sliding simulations starting from $N_{layer}$=7 with either $F_N=0.2$ nN and $F_N=0.9$ nN. 
The simulated system at $F_N=0.2$ nN maintains $7$ layers confined in between the plates,
while that at $F_N=0.9$ nN undergoes a $7 \rightarrow 5$ relayering transition, whose kinetics seems favored by sliding.

As shown in Fig.\ref{fig3}b, this transition corresponds to the sudden drop of the gap $D$, associated with 
the expulsion of the pair of layers, replacing the $N_{layer}$=7 metastable state with the stable $N_{layer}$=5.
This is an example of how the lateral driving can facilitate the squeezout of layers by mechanically perturbing the IL.\\
The close proximity of the $7 \rightarrow 5$ transition can be
predicted by free enthalpy curves, displayed in Fig.\ref{fig3}c. At $F_N=0.9$ nN the energy barrier separating the states with $7$ and $5$
layers drops. The role of sliding appears to be kinetic, facilitating the jump to the more stable $5$ layer state.
%
%
\section{$N_{layer}$- dependent frictional shear stress}
\begin{figure}
\centering
\includegraphics[width=8.5cm,angle=0]{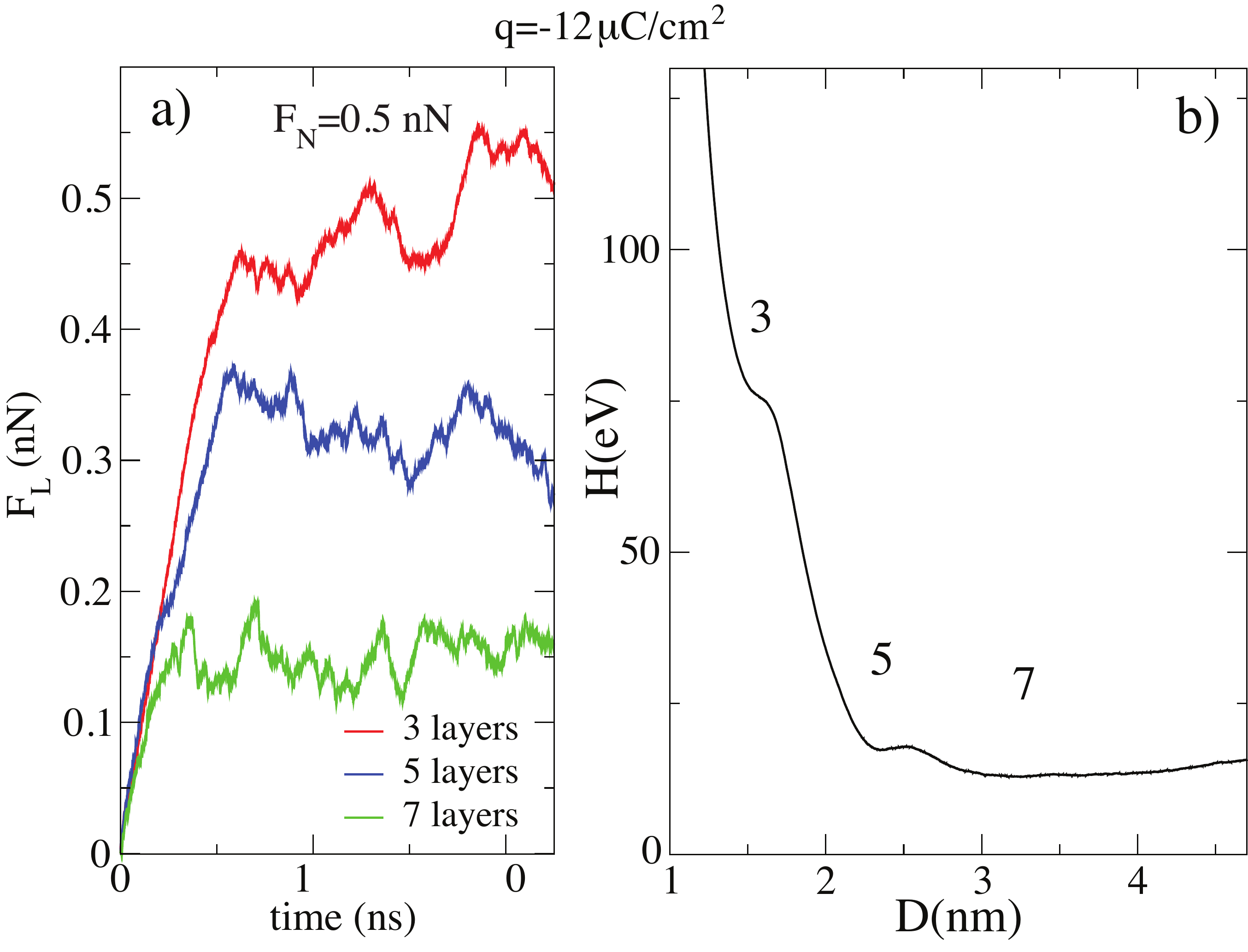}
\caption{a) Lateral force $<F_L>$ as a function of time for plate charge density of $q=-12 \mu C/cm^2$
and for different number of confined layers $N_{layer}$ under the same load $F_N=0.5$. 
b) Static free enthalpy curve for $F_N=0.5$ nN, showing that $N_{layer}$ = 7 is 
stable configuration while 3  and 5 are metastable ones.}
\label{fig4}
\end{figure}

The calculated force-distance curves of Ref.~\cite{capozza2015} show that a single normal load $F_N$ 
generally gives rise to several long lived states with a different number $N_{layer}$. 
For example $F_N=0.5$ nN can support $N_{layer} = 3,5,7$ \cite{capozza2015}, very much as 
it happens in experiment~\cite{hoth14, perkin, perkin10}. 
Figure \ref{fig4}a shows the lateral force $<F_L>$ obtained as a function of time 
for different $N_{layer}$ values, with $F_N=0.5$ nN and $q=-12 \mu C/cm^2$.
Static free enthalpy curve in Fig.\ref{fig4}b demonstrates that $N_{layer}$ = 7 is
stable configuration while 3 and 5 are metastable.
A strong friction increase is observed with decreasing $N_{layer}$, in agreement 
with expectations and with SFA data \cite{perkin,bou-malham10}.
\begin{figure}
\centering
\includegraphics[width=8.5cm,angle=0]{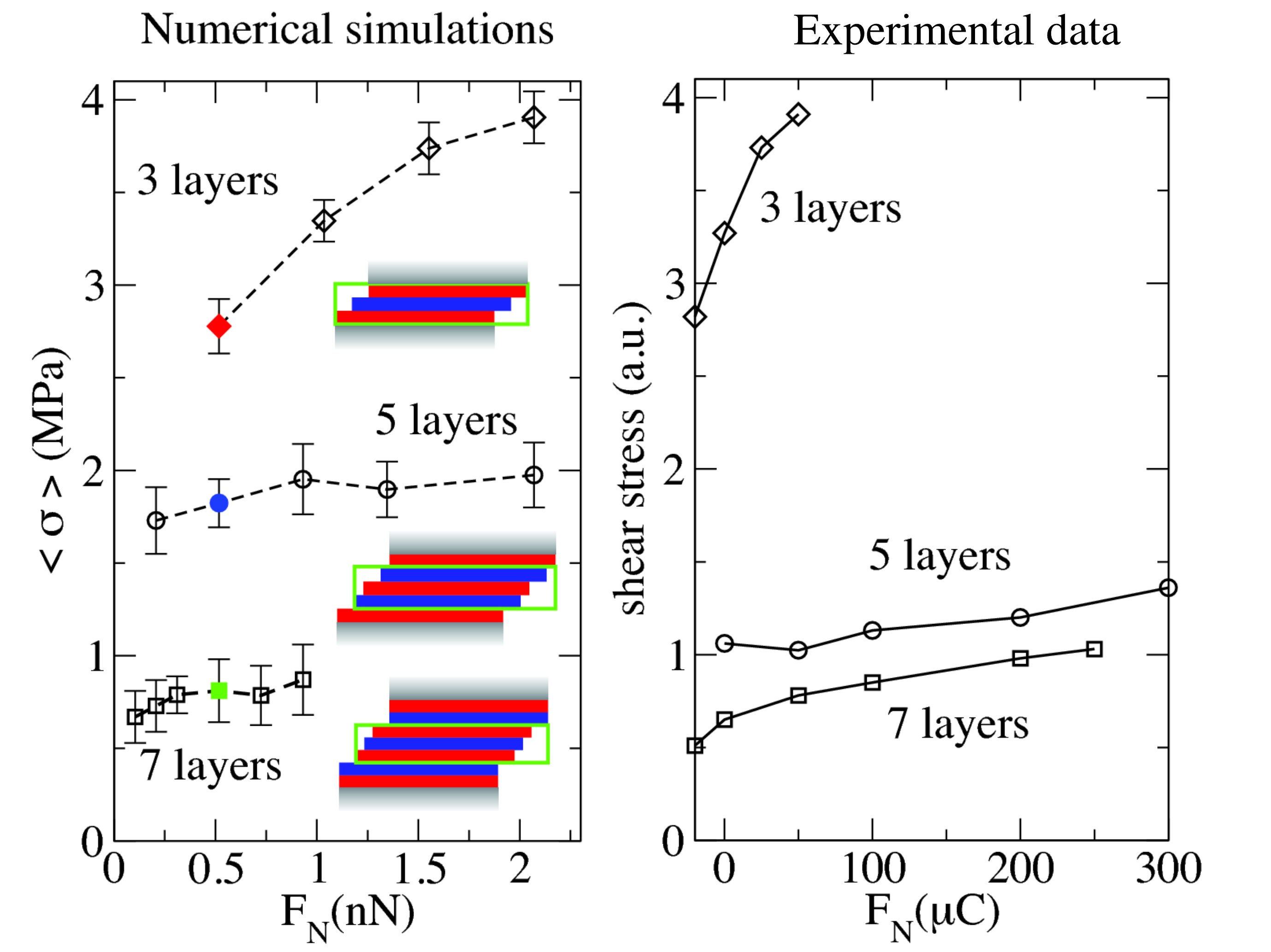}
\caption{a) Average shear frictional stress $<\sigma> = <F_L/A>$ as a function of load $F_N$ for the 
indicated number of confined layers and a charge density of $q=-12 \mu C/cm^2$ on both plates. 
The colored markers refer to the force profiles in panel a) of Fig.\ref{fig4}. 
The cartoons suggest how, upon reducing the number of layers, the shear plane moves closer to interfaces, 
where particles are strongly bound to the plates. b) Approximate behaviour of the shear stress observed with [C4C1Pyrr][NTf2] 
confined between two mica sheets in a Surface Force Balance experiment. This plot has been obtained by rescaling kinetic friction
force data from Ref. \cite{perkin} by a contact area taken as $A\propto {F_T}^{2/3}$ (see text). Theoretical and experimental 
shear stresses agree remarkably well, showing a relatively modest increase with load.}
\label{fig5}
\end{figure}
The detailed load dependence of the frictional shear stress obtained for each fixed number of layers  is more interesting, and is illustrated in Fig.\ref{fig5}a. 
Here the growth of shear stress with increasing load $F_N$ is surprisingly modest. While that  might appear at first sight in disagreement with experiment, 
which show a nearly linear increase of kinetic friction with load \cite{perkin}, we must recall that the contact area is not a constant in SFA,
whereas shear stresses require normalizing to the area $A$. Assuming for SFA two perpendicularly oriented 
elastic cylinders in Hertzian contact, the area  $A \propto {F_T}^{2/3}$, where $F_T=F_{adhesion}+F_N$  is the effective force between the two 
sliders. Extracting an approximate adhesion force by linearly extrapolating the experimental kinetic friction 
to zero in Fig.1 of Ref.~\cite{perkin} , we recover the "experimentally derived" shear stress vs load curves in 
Fig \ref{fig5}b. Remarkably, they now show a weaker experimental dependence on load, 
actually very close to that predicted by simulations in Fig. \ref{fig5}a.\\

\begin{figure}
\centering
\includegraphics[width=8.5cm,angle=0]{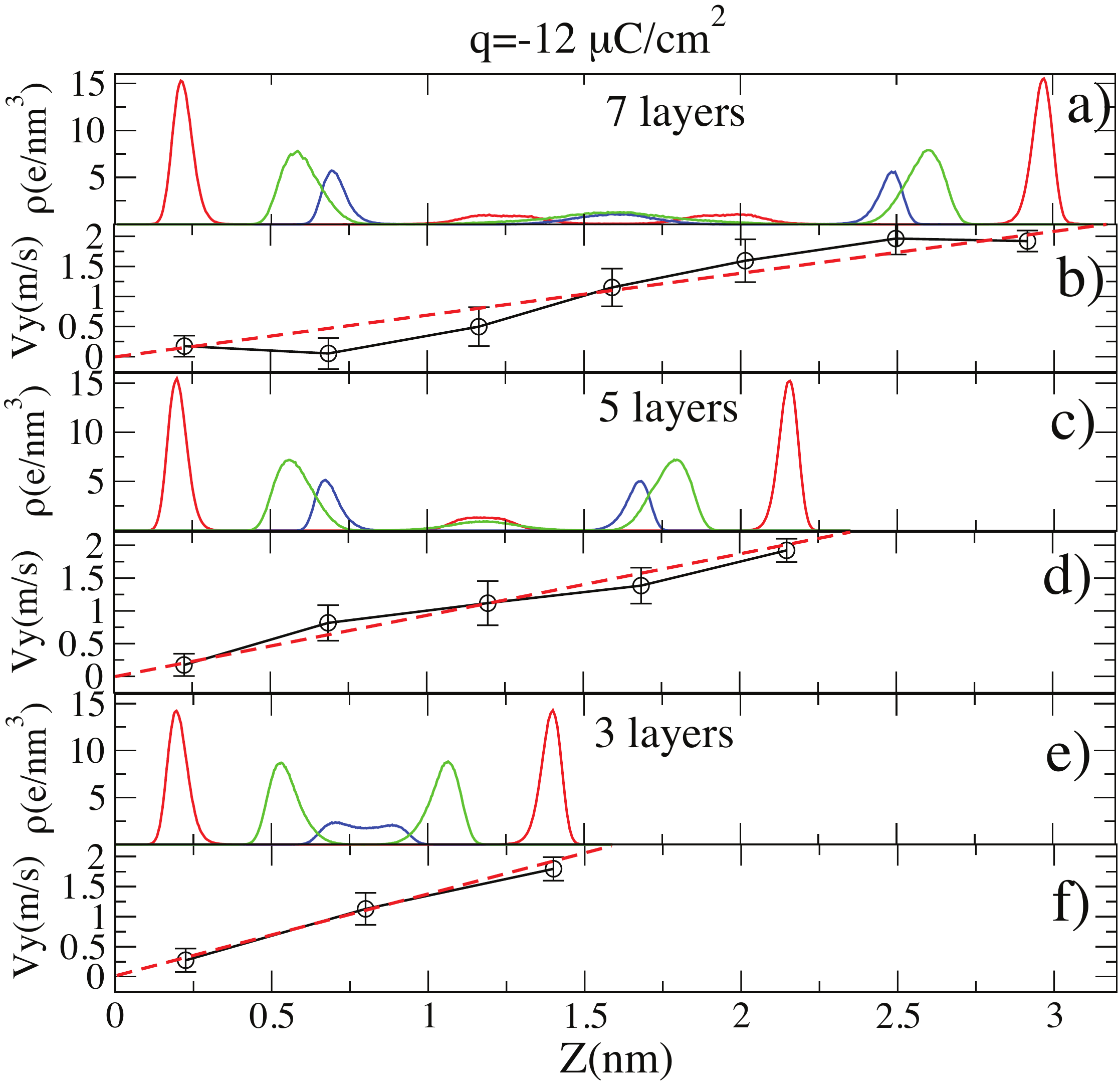}
\caption{Charge density $q=-12 \mu C/cm^2$ on both plates, top plate driven with velocity 
$V=2.2m/s$. Layer-by-layer density profiles ( (a), (c) and (e)) and sliding velocity 
( (b), (d) and (f) ) relative to the bottom plate, for $N_{layer}$=7, $N_{layer}$=5 and 
$N_{layer}$=3 respectively. Red, blue and green curves represents cation, anion and tail 
densities respectively. Comparison with average expected velocities (dashed lines) shows 
that shear is maximal in the three inner layers (the "shear band"), moving closer to the 
plates as $N_{layer}$ decreases. Error bars in panels b, d, f have been reduced ten times
for clarity.}
\label{fig6}
\end{figure}

To understand the strong dependence of $<F_L>$ on the number of layers illustrated in 
Fig.\ref{fig4}a, we now take a closer look to the lubricant shear velocity distribution 
during sliding. Fig.\ref{fig6} shows the layer-by-layer resolved ion density and 
velocity profiles along  $z$  in the liquid for 7, 5 and 3 layers, $F_N=0.5$ nN and 
$V=2.2$ m/s in correspondence of green, blue and red symbols in Fig.\ref{fig5}a respectively.
For $N_{layer}=7$ the velocity profile deviates from the linear behavior (red dashed line) 
indicating that the layers at the boundaries are bound to the plates and tend to move at 
their same velocity. The plate charging promotes the IL in immediate contact to nearly solid 
or at least to a much higher viscosity. Moreover, for this relatively high level of charging, the density 
ratio of first-layer cations over the anode, and of anions over the cathode, is about 1.4, that is  
only slightly larger than one. ~\cite{kornyshev} Both elements support a good grip by the plates. 
The three inner layers constitute by contrast the more liquid, effective shear band where the sliding is
concentrated, and where the coalescence between the two solid-like structures protruding from opposite plates 
is easiest to fracture. 
By reducing the number of layers to 5 and then to 3, the shear band moves closer and closer 
to the boundaries and even the more solid-like structures closer to the plates must break up 
and move, as shown in Fig. \ref{fig6}d, f. As a consequence the lateral force $<F_L>$ 
increases. A pictorial representation of this behavior is provided by the cartoons of 
Fig.\ref{fig6}. 
Surprising at first sight is the seemingly "viscous" friction for $N_{layer}=3$, with a linear 
speed profile distribution between the three layers, the absence of strong stick-slip as 
seen in in Fig.\ref{fig4}a and, as we shall see later in Fig.\ref{fig9}b, a linear friction 
increase with speed. That behavior coexists with a very large shear stress, 
and a very solid structure of this extremely thin lubricant film.
The evolution of layer-by-layer relative velocities shown in Fig.6 indicates that the yield  stresses of
the plate/IL interface and of the mid-film intra-IL interface, which  for $N_{layer}$=7,5 were quite
different, have now become comparable for $N_{layer}$=3.  As a result all interfaces slide by about the same amount. 
The meandering wall $(x,y)$ static structure adopted by the IL in this pressured-and-charged regime ~\cite{capozza2015} 
now incessantly ruptures and rearranges during sliding, leading to a z-uniform velocity gradient,
reminscent of fluid lubrication, despite its nature very different from that of a fluid. 
\begin{figure}
\centering
\includegraphics[width=8.5cm,angle=0]{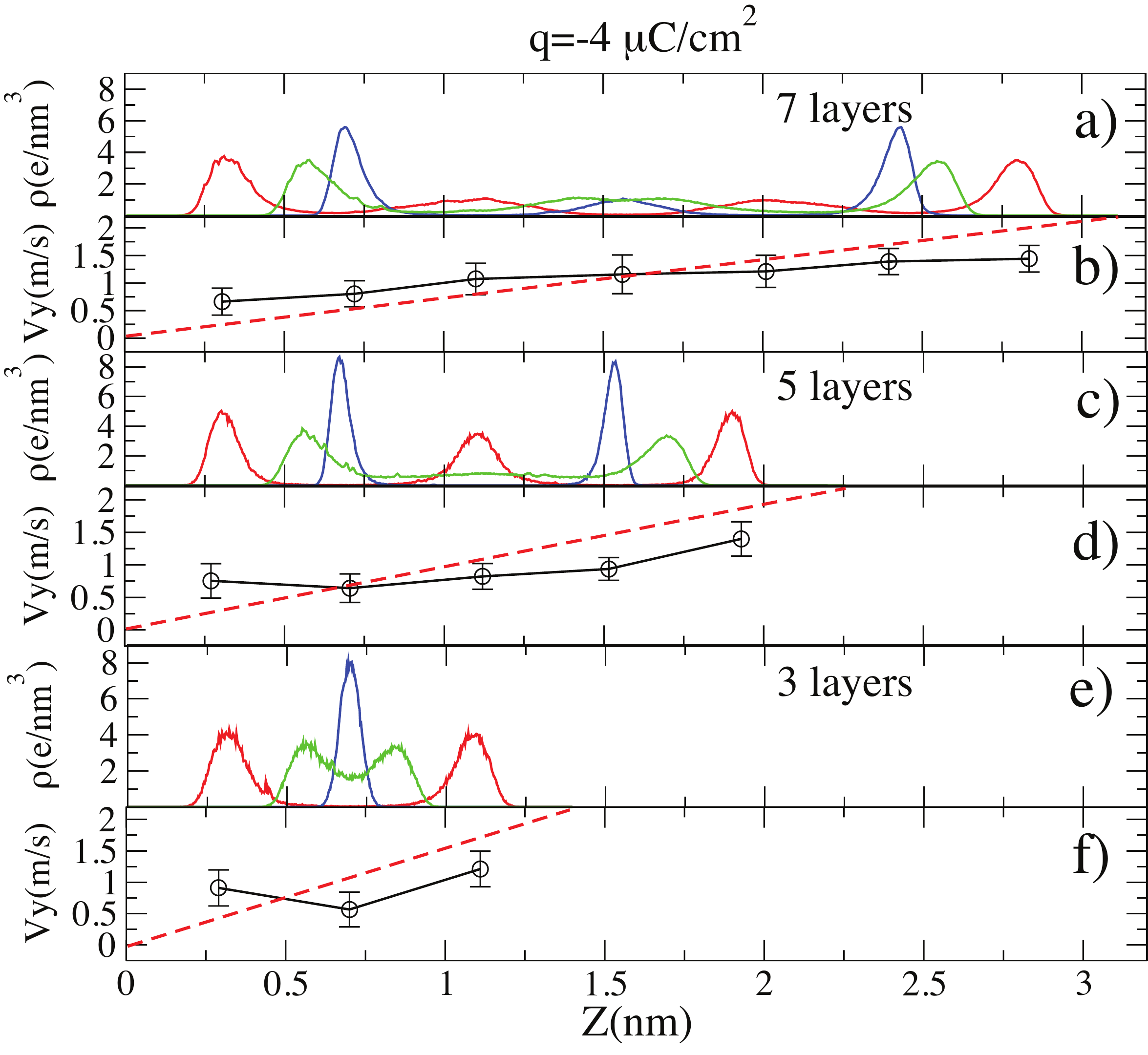}
\caption{Charge density $q=-4 \mu C/cm^2$ on both plates, top plate driven with velocity 
$V=2.2m/s$. Layer-by-layer density profiles (  (a), (c), (e) ) and sliding velocity 
( (b), (d), (f) ) relative to the bottom plate, for 
$N_{layer}$=7, $N_{layer}$=5 and $N_{layer}$=3 respectively. Red, blue and green curves 
represent cation, anion and tail densities respectively.  
For this low charge density the sliding occurs mostly at interface between the IL film 
and the plates. Error bars in panels b, d, f have been reduced ten times
for clarity.}
\label{fig7}
\end{figure}
The picture  changes drastically by reducing the charge to $q=-4\mu C/cm^2$ 
symmetrically on both plates. The mid-film shear band disappears and 
the slippage  occurs mostly at the interfaces between the IL and the plates, 
as shown in Fig.\ref{fig7}. 
While for $q=-12\mu C/cm^2$ the top plate moved by rupturing the solid-like structures reaching 
across plates, at this lower charge the film behaves effectively as a solid
block confined between two slippery surfaces. The shearing concentrates mostly at the two 
interfaces, whence the shear stress depends much more on the plate-IL interaction. 

\begin{figure}
\centering
\includegraphics[width=8.5cm,angle=0]{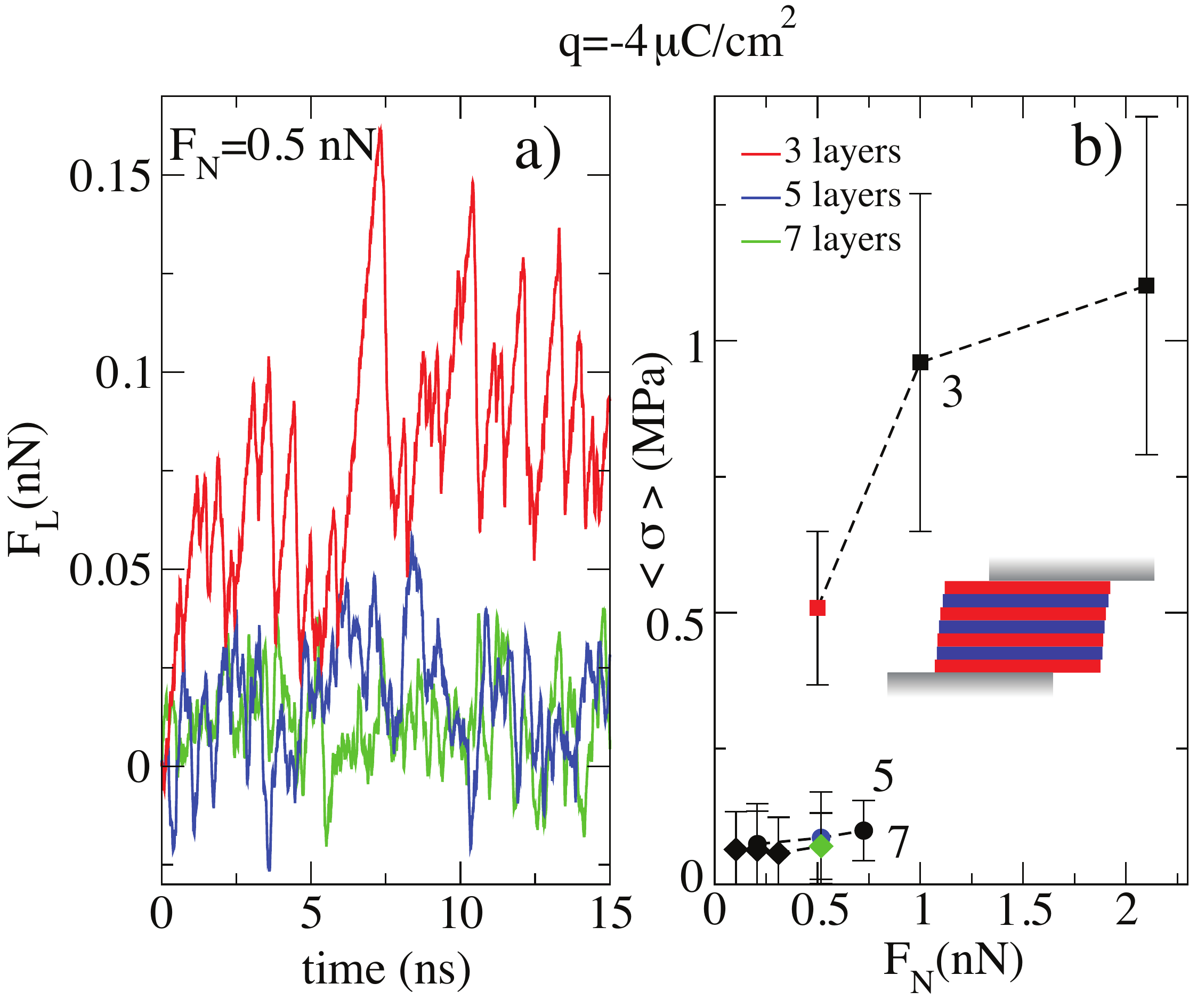}
\caption{a) Frictional force $F_L$ as a function of time for equal and low plate charge density  
$q=-4 \mu C/cm^2$, and for different number of confined layers $N_{layer}$.
b) Average shear stress $\sigma =<F_L>/A$ as a function of $F_N$ for indicated number of confined 
layers and $A=177nm^2$. Colored markers refer to the time profiles shown in panel a).
The cartoon shows that for this low charge density the slippage mostly occurs at
interfaces between the liquid and plates.}
\label{fig8}
\end{figure}

The frictional shear stress $<\sigma>=<F_L>/A$ is about an order of magnitude smaller 
than for higher charge and displays a much weaker dependence on $N_{layer}$, as seen by 
comparing  Fig.\ref{fig8}b with Fig.\ref{fig5}a. 
The rise of average frictional force $<F_L>$ for $N_{layer}=3$, displayed in
Fig.\ref{fig8}a (red line) is associated with a stick-slip regime, as shown by instantaneous 
spring forces in Fig.\ref{fig8}a corresponding to the points indicated in Fig.\ref{fig8}b. 
This is in contrast to the case of high charge, where friction was large but the sliding was smooth. 

The essence for this difference of behavior is explained by the presence, at low but not at large plate charging, 
of overscreening, a phenomenon well demonstrated in earlier work~\cite{kornyshev}. At the small
charging of $|q|= 4\mu C/cm^2$ we find that the charge density ratio of first-layer cations over
that of the plate anode has in fact risen to about 2.3, much larger than one. The overscreened plate
is covered by a dense counterion blanket, making the plate-IL adhesive grip quite effective. The IL film,  
on the other hand, is solid and even harder to fracture. The plate-IL interface yields under sliding and that 
causes stick-slip.\\
\section{Velocity dependence of friction}
\begin{figure}
\centering
\includegraphics[width=8.5cm,angle=0]{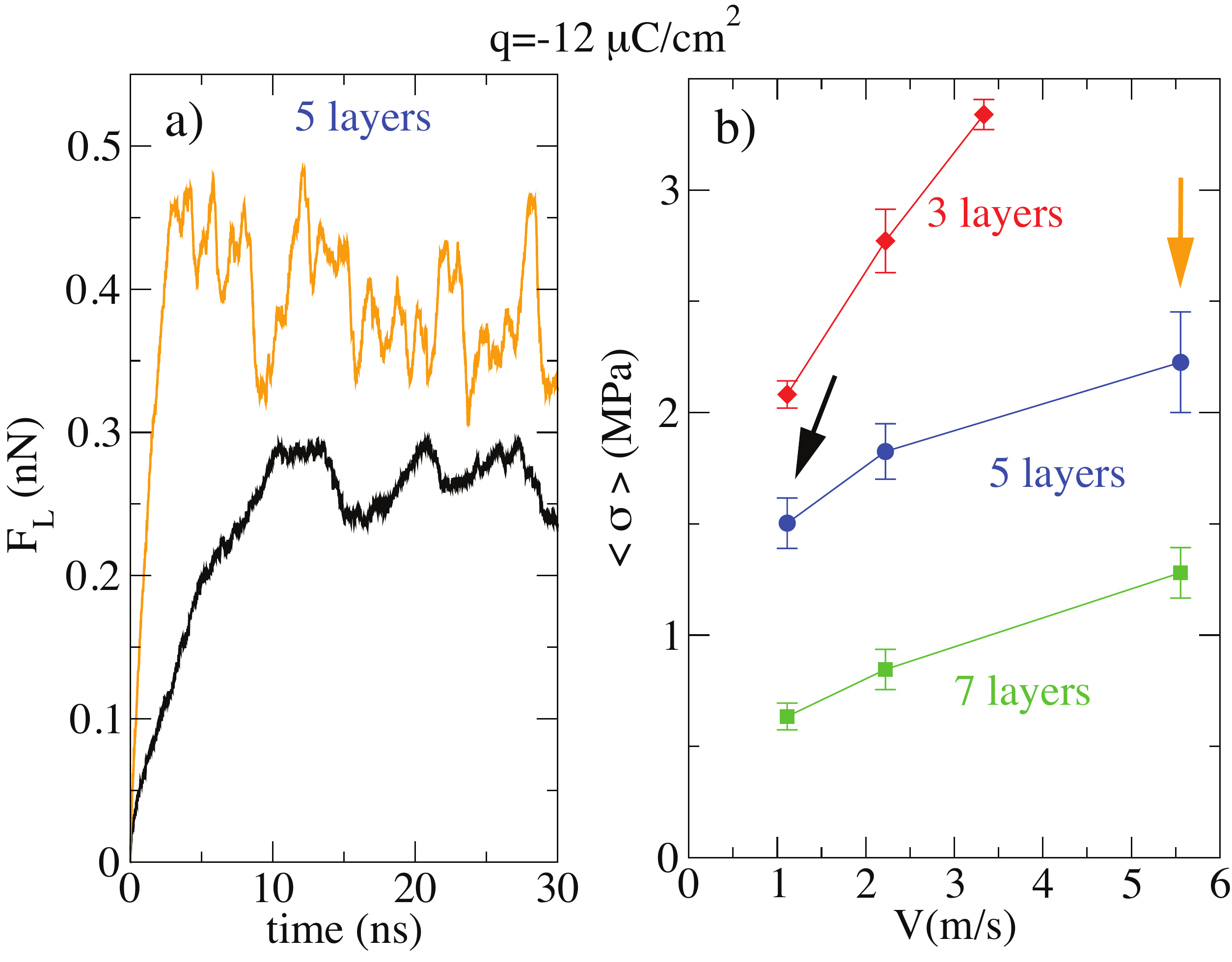}
\caption{a) Frictional force $F_L$ as a function of time for $N_{layer}=5$, $q=-12 \mu C/cm^2$ 
and velocities indicated by arrows in panel b. b) Average shear stress $<\sigma>=<F_L>/A$ as 
a function of driving velocity $V$ for different number of confined layers $N_{layer} =3,5,7$.
The value of normal load here is $F_N=0.5nN$.} 
\label{fig9}
\end{figure}
After the charge, load, and layer number dependence of confined IL friction, we close the 
symmetric plate charging part of this study by investigating the velocity dependence. 
A starting consideration here is that experimental sliding velocities in SFA and AFM are 
very small, typically reaching $10^{-6} m/s$, a value which is at least 6 orders of magnitude 
lower than velocities realistically accessible in atomic and molecular
level simulations. This is a standard difficulty, well known and amply discussed~\cite{vanossi13}; 
one can in fact learn enough even by simulating at very high speeds. Essentially, 
friction can be either smooth and viscous, in which case the shear stress is small 
and proportional to velocity, or it can be stick-slip, in which case the shear stress 
is large and very poorly dependent on velocity. In both cases, very crudely speaking,
even a fast simulation can convey the necessary information (the viscous friction 
coefficient in the first, the full shear stress in the second). Of course it will 
always be necessary to bear in mind that a) increasing velocity may turn 
stick-slip into smooth sliding; and b) microscopic scale stick-slip does not necessarily 
show up as macroscopic or mesoscopic oscillations, and is generally detectable as a 
velocity independent shear stress; c) quantitative aspects 
are controlled by parameters such as the slider's masses, the pulling spring constant, 
etc., whose assumedvalues are largely arbitrary.

\begin{figure}
\centering
\includegraphics[width=8.5cm,angle=0]{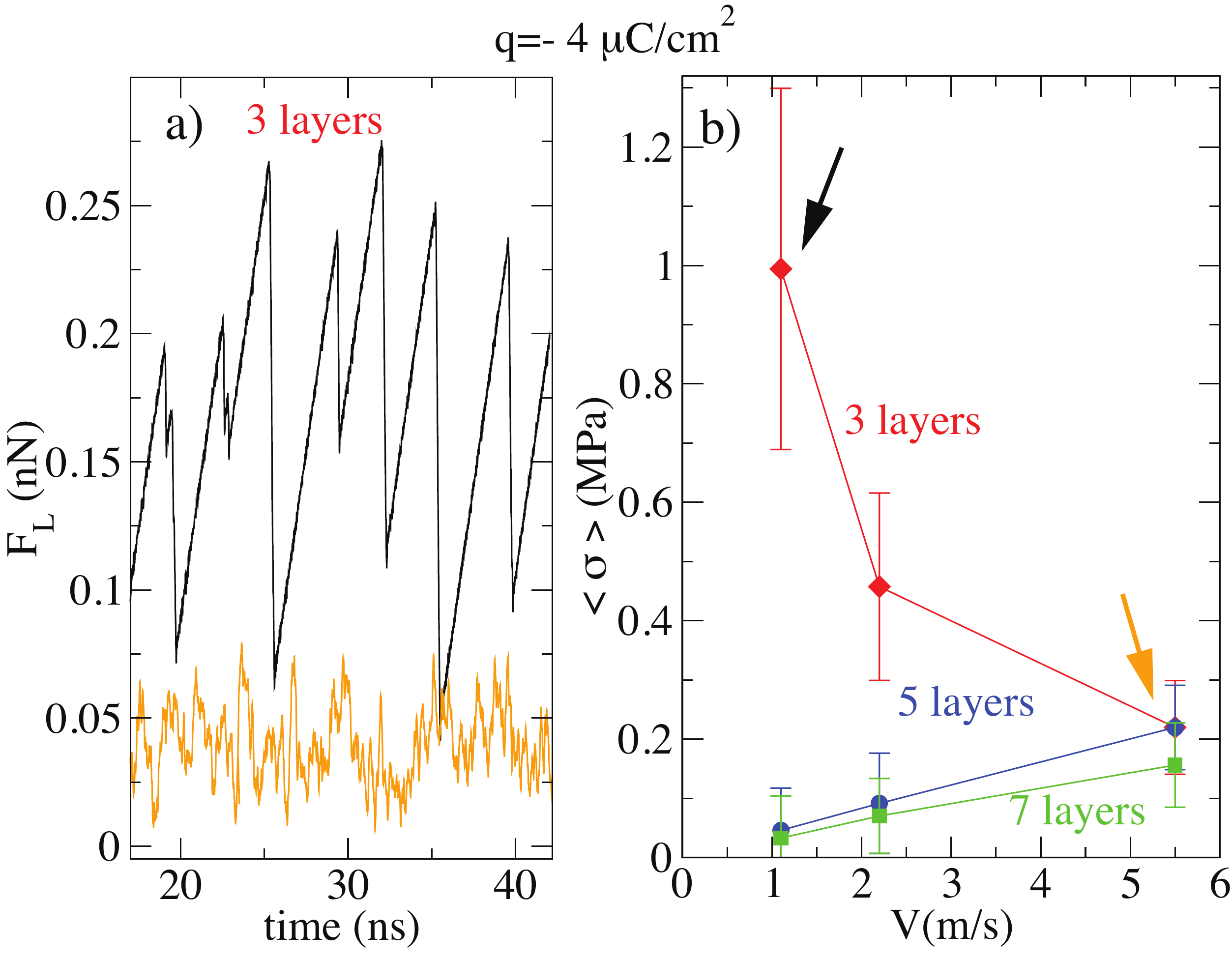}
\caption{a) Lateral force $F_L$ as a function of time for 3 layers, $q=-4 \mu C/cm^2$ 
and velocities indicated by arrows in panel b. b) Average shear frictional stress 
$<\sigma>=<F_L>/A$ as a function of driving velocity $V$ for different number 
of confined layers $N_{layer}$. The value of normal load here is $F_N=0.5nN$.}
\label{fig10}
\end{figure}

With these provisions we investigate the dependence of IL friction upon driving velocity $V$, 
at fixed normal load $F_N=0.5nN$. Fig.\ref{fig9} and \ref{fig10} show the frictional shear 
stress as a function of $V$ for equal-sign charge densities of $q=-12 \mu C/cm^2$ 
and $q=-4 \mu C/cm^2$ respectively and for $N_{layer} =3,5,7$. 
All results for $q=-12 \mu C/cm^2$ in Fig.\ref{fig9} show a linear increase of shear 
stress with velocity indicating a viscous-type friction. As suggested earlier, this is most 
likely attributable to overscreening of both plates by the extra anions. However 
$<\sigma>$ vs $V$ does not extrapolate to $0$, so that even at $V\simeq 0$ a finite force 
is needed to set the system in motion. This is the force needed in order to rupture 
the vertical IL "solid" walls that straddle across the two charged plates (Fig.\ref{fig5}).\\

Also for low charge density $q=-4 \mu C/cm^2$, the shear stress for $N_{layer} =5,7$ 
grows proportional to velocity but here it extrapolates to zero as $V \to 0$ (see Fig.\ref{fig10}b). 
This is the hallmark of true viscous friction, and it agrees with the top plate moving in a 
clear regime of smooth sliding. The sliding regime drastically changes for $N_{layer} =3$  
and low velocity $V=1.1m/s$, where a clear stick-slip regime takes over. The alternation 
of sticking periods followed  by rapid and highly dissipative motion 
of the top plate is responsible for the strong increase of average lateral force 
at low velocities, as demonstrated in Fig.\ref{fig10}a. By increasing $V$, the stick-slip 
first turns chaotic and eventually disappears at $V=5.5m/s$, where smooth sliding is recovered.
Particle trajectories clearly show this transition. At low $V$ the IL structure retains
its worm-like meandering planes or chains, which we described in ref. \cite{capozza2015}, and
which were recently observed in experiments, ~\cite{segura13} 
and behaves essentially as a solid confined between two slippery planes. 

The transition from stick-slip to smooth sliding is shown in Fig.\ref{fig10}a, reporting two
force profiles for $N_{layer} =3$ corresponding to driving velocities indicated by arrows 
in Fig.\ref{fig10}b.
\section{Friction with opposite plate charging}

In this final part we study how friction force is affected by opposite charging of the two
confining plates, such as one would obtain by applying a DC voltage between them. 
Probably not that easy to realize experimentally, this configuration is nonetheless 
quite interesting.
Upon balancing the opposite charge on the two plates, the equilibrium layering structure
now forms an even number of layers, a rearrangement which preserves local charge neutrality. 
In order to explore expeditely the dependence of friction upon the magnitude of the 
charge imbalance, we conduct a time-dependent simulation where we periodically, 
but very slowly and as adiabatically as possible, modulate in time the on-plate charge 
in the form $q = \pm Q \sin (2\pi t/\tau)$, where the two opposite signs refer to the 
two opposite plates. Adiabaticity is pursued by choosing a period $\tau$ large enough 
compared to the characteristic time scale of all relevant ionic rearrangements. 

The magnitude of charging used in this work, $Q \leq 12 \mu C/cm^2$ is dictated, as discussed previously
~\cite{capozza2015}, by the requirement that, assuming an average dielectric constant of 2, the plate-plate
voltage should not exceed 5V, taken as a reasonable upper limit in an experimental situation.
In this respect therefore, our study does not overlap with previous work, where in a different model \cite{fajardo15} 
the behavior at $Q \geq 16 \mu C/cm^2$ was investigated. 

\begin{figure}
\centering
\includegraphics[width=8.5cm,angle=0]{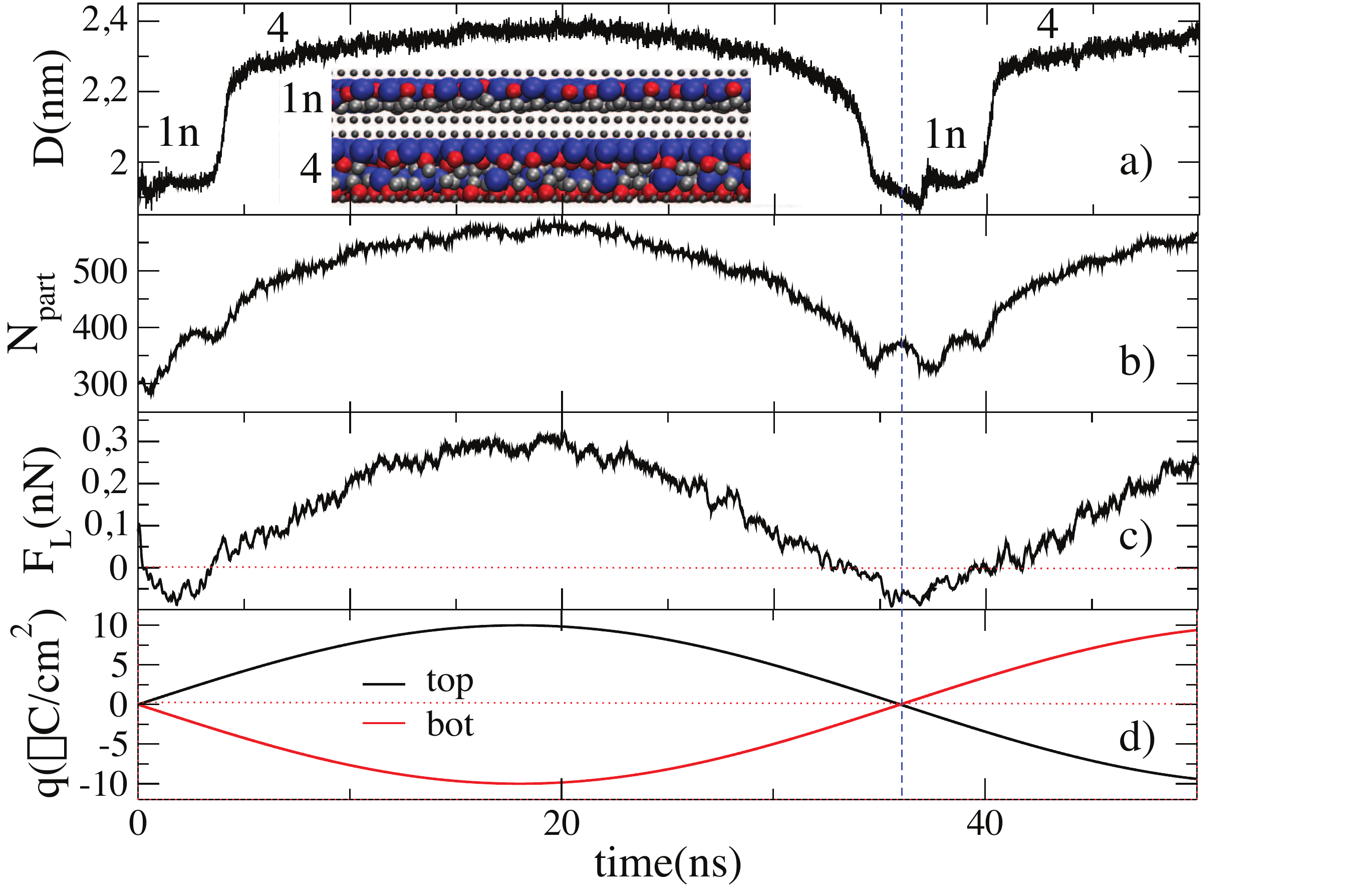}
\caption{a) 
Frictional sliding evolution while slowly changing opposite charges on the two sliding plates. 
Parameters: Initial  $N_{layer} =1$, $F_N= 1$ nN,  charges  $q = \pm Q \sin (2\pi t/\tau)$
with $Q = 10\mu C/cm^2$ and  $\tau = 72$ ns, sliding velocity $V= 22m/s$.
a) Spontaneous evolution of plate-plate distance $D$. 
b) Number of particles $N_{part}$ in the interplate gap. 
c) Sliding  force $F_L$ as a function of time.
d) Time evolution of top and bottom plate charge. 
Charging causes the plate separation $D$ to open up at charge maxima (IL suck-in) and to 
collapse at charge zeroes (IL squeezout), all under constant load. The inset in panel a)
shows a side view of the confined liquid structure for 4 layers  
($q = Q$) and for the single neutral planar layer $1n$ ($q = 0$).
Friction is large when $|q|$ and voltage are maxima, and charging solidifies the IL inside 
the gap $D$. Conversely, when $q$ is near zero, the IL melts, the plates collapse together 
squeezing out all fluid except for a neutral monolayer where both ions pack in a square lattice, 
resulting in a very low friction.}
\label{fig11}
\end{figure}

\begin{figure}
\centering
\includegraphics[width=8.5cm,angle=0]{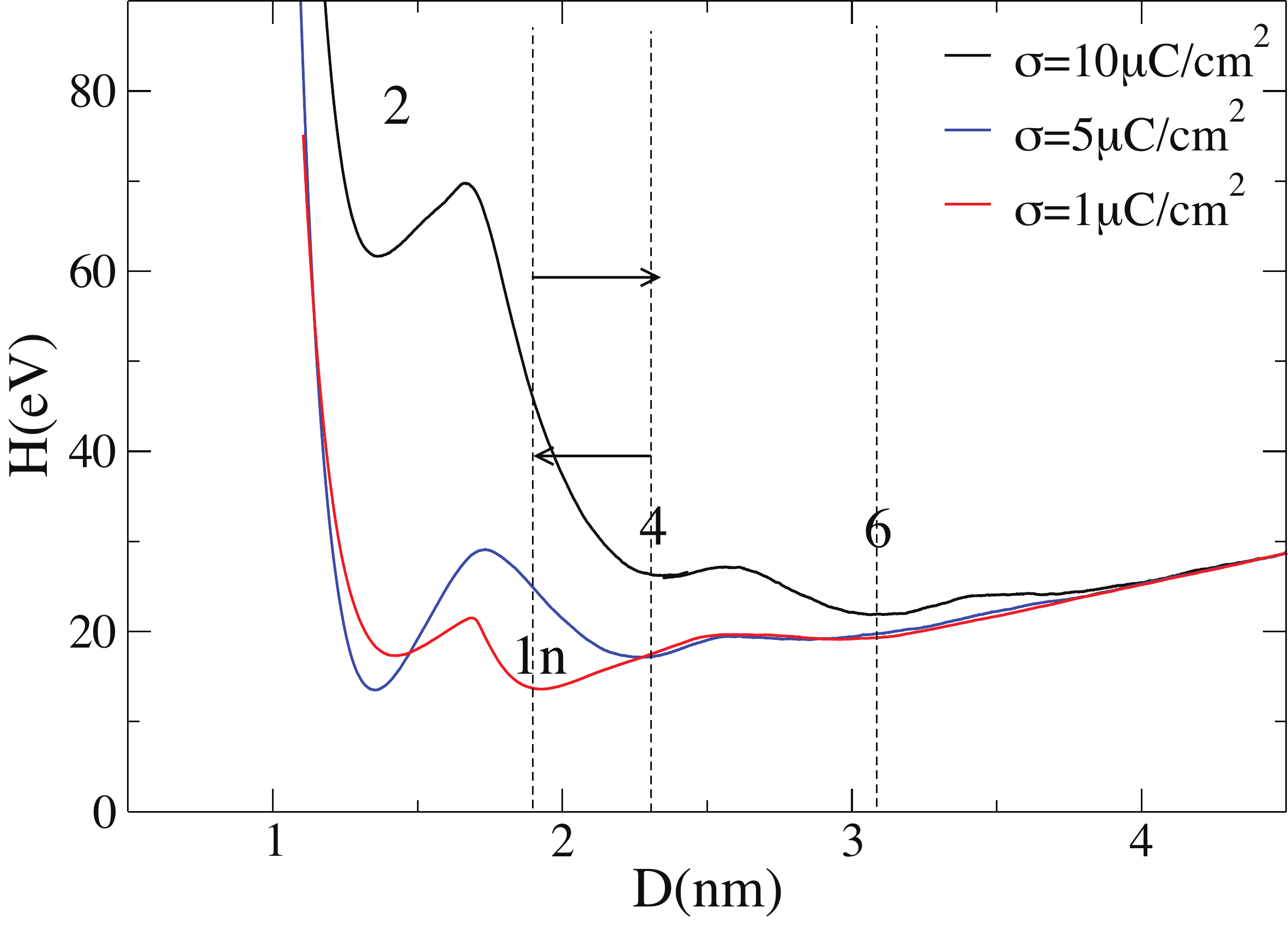}
\caption{(Color online) Enthalpy curves for indicated values of charge
density on plates. Numbers indicated in figure denote metastable states at
different number of layers. Peak heights decrease with charge
and arrows indicate charge induced transitions between the states $1n$ and $4$.}
\label{fig12}
\end{figure}

Assuming a constant load  $F_N = 1$ nN, a charge oscillation period $\tau = 72$ ns, 
a magnitude $Q = 10\mu C/cm^2$ and a sliding velocity $V=22m/s$
we obtain the spontaneous evolution of the system shown in Fig.\ref{fig11}. What happens is,
in a nutshell, that the interplate distance $D$ spontaneously changes, and so does 
the sliding friction $F_L$ as the charge changes.
The simulation begins with a charge density of $|q(t = 0)| = 0$ on plates and the IL 
arranges in one single neutral layer indicated as $1n$ in Fig.\ref{fig11}, which, as 
shown in the static squeezout enthalpy of Fig.\ref{fig12}, is the most stable configuration 
under the fixed applied loading force $F_N = 1$ nN and low charge. 
As the plate charge increases, the liquid in the lateral reservoirs is sucked in 
the gap (despite a constant load), reaching a four layer solid state. 
Actually, we expected it to go back to six layers, which is the
equilibrium free enthalpy minimum, as shown in Fig.\ref{fig12}. It probably would 
if we could simulate for much longer times, but it does not on our limited time scale, 
due to tha strongly reduced mobility of ions in the near-solid film.
This strong and rigid four layers structure softens as the plate charge decreases, 
eventually turning more fluid and squeezing out reaching again the single neutral layer
arrangement $1n$. These charging-induced relayering and squeezout are accompanied by 
drastic change in frictional behaviour of the liquid film. The large shear stress at 
high plate charge (high voltage, near 5 eV) is replaced by a nearly vanishing shear 
stress at zero charge. Here therefore, contrary to the equal-charge case of the previous 
sections, the thinnest film yields the lowest friction, rather than the opposite. 
This is because, independently of sliding, the plate charging induces structural 
transformations of freezing/melting of the the IL film\cite{capozza2015}.
The frozen IL supports the applied load for large charge but the liquid 
IL does not for zero charge. The IL
film structure at $q\simeq 0$ shrinks to a single, planar, square and neutral IL monolayer. 
Here the neutral tails stick out of the plane, effectively lubricating the sliding. 
A qualitatively similar mechanism where alkyl tails stick out of a well-packed anion-cation 
plane has been proposed in a recent experimental study \cite{bennewitz} 
to explain the potential-dependent frictional behaviour of an AFM tip in presence of an IL.\\
We did change and explore a variety of parameters, including charging magnitude, 
applied load, and sliding speed in order to test the main aspects
of these results. The processes of melting and freezing, and of expulsion and sucking always remained very asymmetrical between the 
fast squeezout, associated with sudden melting of the IL film upon decreasing $|q|$ towards zero, and the more gradual 
sucking-in and freezing, attained upon increasing $|q|$ from zero in Fig.\ref{fig11}a. The electrically driven fast emptying 
but slow refilling of the interplate gap could be used, together with some spatial asymmetry to build an effective effective electro-pumping
of ions in and out of the gap. 
\begin{figure}
\centering
\includegraphics[width=8.5cm,angle=0]{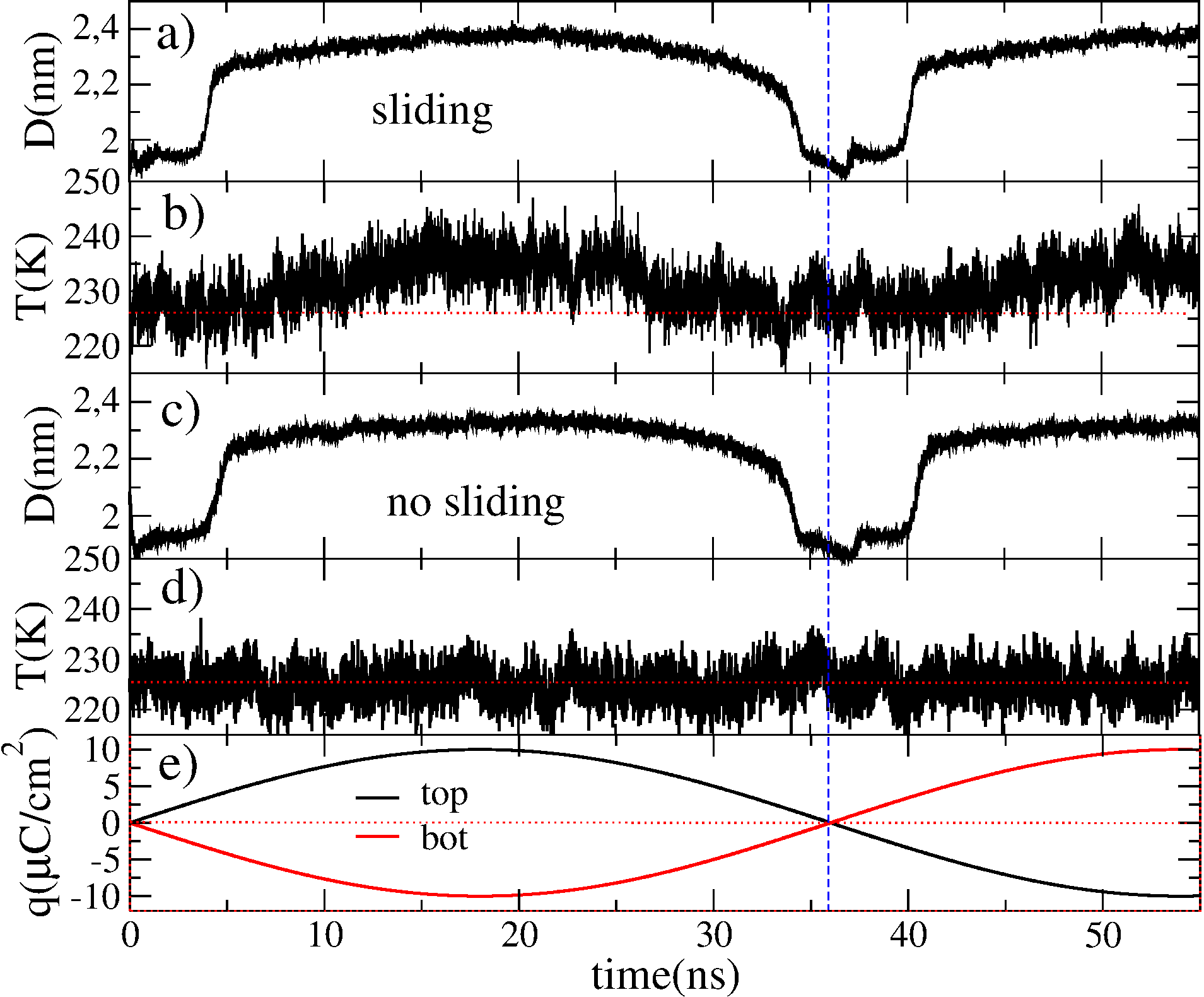}
\caption{(Color online) Comparison between the charge induced squeezout and relayering
processes in the presence and absence of sliding. 
Spontaneous evolution of plate-plate distance $D$ ((a) and (c)) and temperature 
((b) and (d)) in comparison with time evolution of top and bottom plate charge (e).
In the absence of driving the temperature remains stable at the set value of $T=225K$, 
while in the presence of driving it temporarily deviates from this value.
}
\label{fig13}
\end{figure}
The frictional shear stress shows a very large charge dependence, as a direct consequence of the changes of
structure and of gap width $D(q)$ caused by charging. Although noisy, the large shear stress at charge maxima
drops by an order of magnitude or more following squeezout at zero charge. This strong effect appears promising
in view of experimental verifications, and of a possible use for the control of friction.  

Lastly, we briefly dwell on questions connected with work, heat and temperature, which come about with respect both to 
charging-induced melting and freezing, and to frictional work and heating. In the absence of sliding, electrically
induced freezing and melting would be associated with the absorption or release of latent heat. That would be large if the solid was 
ordered, e.g., crystalline or nearly crystalline, as is the case for example in the NaCl model of the IL ~\cite{capozza2015}.
The more the solid resembles a glass, however, the smaller the latent heat will be, owing to structural near-identity
between glass and liquid. That is the case of our TM model, where indeed the latent heat of melting was very small in
the simulated caloric curves~\cite{capozza2015}. Without sliding, the presence of a latent heat is undetectable in electrically
induced freezing and melting cycles shown in Fig.\ref{fig13}c,d. Of course, these data were obtained with a thermostat, canceling
any large temperature fluctuations: despite the themostat, dips and peaks would be well visible in presence of a robust,
crystal-like latent heat.    
Frictional heating is instead quite visible in the sliding results under cycled charging of Fig.\ref{fig13}a,b. 
Despite the thermostat, the temperature
is seen to oscillate, downwards when  the solid confined IL lubricant goes liquid at $q=0$, up when it goes back to solid at $|q|= maximum$ 
and where therefore the frictional Joule heating is much stronger, down again upon melting where the friction drops once again. 
Because of the presence of the thermostat we did not attempt a quantitative connection between the frictional work and the
residual temperature oscillation, but the connection between the two is quite clear. 

\section{Conclusions}

The electrical charging of plates in nearly atomic contact influences an IL when trapped in a nanoscale sized gap
between them. As a result, the frictional shear stress realized upon mutual sliding of the plates is modified by the charging. The present 
simulation study conducted within the simple "tailed model" previously developed for the IL explores some of the rich variety 
of phenomena that can take place in this context. The IL becomes structured into alternating charge layers, whose number
is odd when the plates are equally charged, even when they are oppositely charged. The trapped IL film may develop a solid-like
rigidity, with different and opposite frictional consequences depending on charging level and other parameters. When the plate
grips the solid-like film and the shearing takes place in its middle the friction is large; but when the grip is less effective the shear
concentrates at the plate-film interface and friction drops. Alternatively, at zero charging and under the same conditions the ionic 
liquid can melt, thus lubricating and dramatically reducing the sliding friction.

Even though the parameter-dependent nature of our results does not make a comparison with any specific experimental system 
particularly compelling, we believe that several elements uncovered either connect with observations already
available, or else suggest pursuing newer ones.  The modest dependence of friction upon load -- unusual in the context
of classical friction -- is among the former. The possibility that electrical charging, in whichever manners that could be realized, 
could influence the tribology of confined IL by causing its electrically driven freezing and melting is among the latter, and 
deserves to be tested, supporting the possibility of an electric control of IL lubricated friction.

\section{acknowledgments}
The authors are grateful to Daniele Passerone and Carlo Pignedoli from EMPA (D\"ubendorf, CH) 
for the computational resources and the technical assistance provided, and, at various times, for discussions 
with A. Kornyshev and A. Schirmeisen. Work in Trieste was partly sponsored 
by Sinergia Contract CRSII2$_1$36287/1, and by ERC Advanced Grant 320796 - MODPHYSFRICT. This work 
is also supported by the COST Action MP1303 "Understanding and Controlling Nano and Mesoscale Friction".

\bibliography{biblio}

\begin{thebibliography}{34}
\expandafter\ifx\csname natexlab\endcsname\relax\def\natexlab#1{#1}\fi
\expandafter\ifx\csname bibnamefont\endcsname\relax
  \def\bibnamefont#1{#1}\fi
\expandafter\ifx\csname bibfnamefont\endcsname\relax
  \def\bibfnamefont#1{#1}\fi
\expandafter\ifx\csname citenamefont\endcsname\relax
  \def\citenamefont#1{#1}\fi
\expandafter\ifx\csname url\endcsname\relax
  \def\url#1{\texttt{#1}}\fi
\expandafter\ifx\csname urlprefix\endcsname\relax\def\urlprefix{URL }\fi
\providecommand{\bibinfo}[2]{#2}
\providecommand{\eprint}[2][]{\url{#2}}

\bibitem[{\citenamefont{Plechkova and Seddon}(2008)}]{plechkova08}
\bibinfo{author}{\bibfnamefont{N.~V.} \bibnamefont{Plechkova}}
  \bibnamefont{and} \bibinfo{author}{\bibfnamefont{K.~R.}
  \bibnamefont{Seddon}}, \bibinfo{journal}{Chem. Soc. Rev.}
  \textbf{\bibinfo{volume}{37}}, \bibinfo{pages}{123} (\bibinfo{year}{2008}).

\bibitem[{\citenamefont{Hallett and Welton}(2011)}]{hallett11}
\bibinfo{author}{\bibfnamefont{J.~P.} \bibnamefont{Hallett}} \bibnamefont{and}
  \bibinfo{author}{\bibfnamefont{T.}~\bibnamefont{Welton}},
  \bibinfo{journal}{Chem. Rev.} \textbf{\bibinfo{volume}{111}},
  \bibinfo{pages}{3508} (\bibinfo{year}{2011}).

\bibitem[{\citenamefont{Hayes et~al.}(2010)\citenamefont{Hayes, Warr, and
  Atkin}}]{hayes10}
\bibinfo{author}{\bibfnamefont{R.}~\bibnamefont{Hayes}},
  \bibinfo{author}{\bibfnamefont{G.~G.} \bibnamefont{Warr}}, \bibnamefont{and}
  \bibinfo{author}{\bibfnamefont{R.}~\bibnamefont{Atkin}},
  \bibinfo{journal}{Phys. Chem. Chem. Phys.} \textbf{\bibinfo{volume}{12}},
  \bibinfo{pages}{1709} (\bibinfo{year}{2010}).

\bibitem[{\citenamefont{Perkin et~al.}(2010)\citenamefont{Perkin, Albrecht, and
  Klein}}]{perkin10}
\bibinfo{author}{\bibfnamefont{S.}~\bibnamefont{Perkin}},
  \bibinfo{author}{\bibfnamefont{T.}~\bibnamefont{Albrecht}}, \bibnamefont{and}
  \bibinfo{author}{\bibfnamefont{J.}~\bibnamefont{Klein}},
  \bibinfo{journal}{Phys. Chem. Chem. Phys.} \textbf{\bibinfo{volume}{12}},
  \bibinfo{pages}{1243} (\bibinfo{year}{2010}).

\bibitem[{\citenamefont{Espinosa-Marzal
  et~al.}(2014)\citenamefont{Espinosa-Marzal, Arcifa, Rossi, and
  Spencer}}]{spencer}
\bibinfo{author}{\bibfnamefont{R.~M.} \bibnamefont{Espinosa-Marzal}},
  \bibinfo{author}{\bibfnamefont{A.}~\bibnamefont{Arcifa}},
  \bibinfo{author}{\bibfnamefont{A.}~\bibnamefont{Rossi}}, \bibnamefont{and}
  \bibinfo{author}{\bibfnamefont{N.~D.} \bibnamefont{Spencer}},
  \bibinfo{journal}{J. Phys. Chem. Lett.} \textbf{\bibinfo{volume}{5}},
  \bibinfo{pages}{179} (\bibinfo{year}{2014}).

\bibitem[{\citenamefont{Werzer et~al.}(2012)\citenamefont{Werzer, Cranston,
  Warr, Atkin, and Rutland}}]{werzer12}
\bibinfo{author}{\bibfnamefont{O.}~\bibnamefont{Werzer}},
  \bibinfo{author}{\bibfnamefont{E.~D.} \bibnamefont{Cranston}},
  \bibinfo{author}{\bibfnamefont{G.~G.} \bibnamefont{Warr}},
  \bibinfo{author}{\bibfnamefont{R.}~\bibnamefont{Atkin}}, \bibnamefont{and}
  \bibinfo{author}{\bibfnamefont{M.}~\bibnamefont{Rutland}},
  \bibinfo{journal}{Phys. Chem. Chem. Phys.} \textbf{\bibinfo{volume}{14}},
  \bibinfo{pages}{5147} (\bibinfo{year}{2012}).

\bibitem[{\citenamefont{Li et~al.}(2013{\natexlab{a}})\citenamefont{Li,
  Rutland, and Atkin}}]{li13}
\bibinfo{author}{\bibfnamefont{H.}~\bibnamefont{Li}},
  \bibinfo{author}{\bibfnamefont{M.~W.} \bibnamefont{Rutland}},
  \bibnamefont{and} \bibinfo{author}{\bibfnamefont{R.}~\bibnamefont{Atkin}},
  \bibinfo{journal}{Phys. Chem. Chem. Phys.} \textbf{\bibinfo{volume}{15}},
  \bibinfo{pages}{14616} (\bibinfo{year}{2013}{\natexlab{a}}).

\bibitem[{\citenamefont{Sweeney et~al.}(2012)\citenamefont{Sweeney, Hausen,
  Hayes, Webber, Endres, Rutland, Bennewitz, and Atkin}}]{bennewitz}
\bibinfo{author}{\bibfnamefont{J.}~\bibnamefont{Sweeney}},
  \bibinfo{author}{\bibfnamefont{F.}~\bibnamefont{Hausen}},
  \bibinfo{author}{\bibfnamefont{R.}~\bibnamefont{Hayes}},
  \bibinfo{author}{\bibfnamefont{G.~B.} \bibnamefont{Webber}},
  \bibinfo{author}{\bibfnamefont{F.}~\bibnamefont{Endres}},
  \bibinfo{author}{\bibfnamefont{M.~W.} \bibnamefont{Rutland}},
  \bibinfo{author}{\bibfnamefont{R.}~\bibnamefont{Bennewitz}},
  \bibnamefont{and} \bibinfo{author}{\bibfnamefont{R.}~\bibnamefont{Atkin}},
  \bibinfo{journal}{Phys. Rev. Lett.} \textbf{\bibinfo{volume}{109}},
  \bibinfo{pages}{155502} (\bibinfo{year}{2012}).

\bibitem[{\citenamefont{Li et~al.}(2014)\citenamefont{Li, Wood, Rutland, and
  Atkin}}]{li14}
\bibinfo{author}{\bibfnamefont{H.}~\bibnamefont{Li}},
  \bibinfo{author}{\bibfnamefont{R.~J.} \bibnamefont{Wood}},
  \bibinfo{author}{\bibfnamefont{M.~W.} \bibnamefont{Rutland}},
  \bibnamefont{and} \bibinfo{author}{\bibfnamefont{R.}~\bibnamefont{Atkin}},
  \bibinfo{journal}{Chem. Commun.} \textbf{\bibinfo{volume}{50}},
  \bibinfo{pages}{4368} (\bibinfo{year}{2014}).

\bibitem[{\citenamefont{Hoth et~al.}(2014)\citenamefont{Hoth, Hausen,
  M\"{u}ser, and Bennewitz}}]{hoth14}
\bibinfo{author}{\bibfnamefont{J.}~\bibnamefont{Hoth}},
  \bibinfo{author}{\bibfnamefont{F.}~\bibnamefont{Hausen}},
  \bibinfo{author}{\bibfnamefont{M.~H.} \bibnamefont{M\"{u}ser}},
  \bibnamefont{and}
  \bibinfo{author}{\bibfnamefont{R.}~\bibnamefont{Bennewitz}},
  \bibinfo{journal}{J. Phys.: Condens. Matter} \textbf{\bibinfo{volume}{26}},
  \bibinfo{pages}{284110} (\bibinfo{year}{2014}).

\bibitem[{\citenamefont{Mendonc\c{a} et~al.}(2013)\citenamefont{Mendonc\c{a},
  P\`{a}dua, and Malfreyt}}]{padua13}
\bibinfo{author}{\bibfnamefont{A.}~\bibnamefont{Mendonc\c{a}}},
  \bibinfo{author}{\bibfnamefont{A.}~\bibnamefont{P\`{a}dua}},
  \bibnamefont{and} \bibinfo{author}{\bibfnamefont{P.}~\bibnamefont{Malfreyt}},
  \bibinfo{journal}{J. Chem. Theory Comput.} \textbf{\bibinfo{volume}{9}},
  \bibinfo{pages}{1600} (\bibinfo{year}{2013}).

\bibitem[{\citenamefont{Federici-Canova
  et~al.}(2014)\citenamefont{Federici-Canova, Matsubara, Mizukami, Kurihara,
  and Shluger}}]{federici14}
\bibinfo{author}{\bibfnamefont{F.}~\bibnamefont{Federici-Canova}},
  \bibinfo{author}{\bibfnamefont{H.}~\bibnamefont{Matsubara}},
  \bibinfo{author}{\bibfnamefont{M.}~\bibnamefont{Mizukami}},
  \bibinfo{author}{\bibfnamefont{K.}~\bibnamefont{Kurihara}}, \bibnamefont{and}
  \bibinfo{author}{\bibfnamefont{A.~L.} \bibnamefont{Shluger}},
  \bibinfo{journal}{Phys. Chem. Chem. Phys.} \textbf{\bibinfo{volume}{16}},
  \bibinfo{pages}{8247} (\bibinfo{year}{2014}).

\bibitem[{\citenamefont{Fajardo et~al.}(2015)\citenamefont{Fajardo, Bresme,
  Kornyshev, and Urbakh}}]{fajardo15}
\bibinfo{author}{\bibfnamefont{O.}~\bibnamefont{Fajardo}},
  \bibinfo{author}{\bibfnamefont{F.}~\bibnamefont{Bresme}},
  \bibinfo{author}{\bibfnamefont{A.}~\bibnamefont{Kornyshev}},
  \bibnamefont{and} \bibinfo{author}{\bibfnamefont{M.}~\bibnamefont{Urbakh}},
  \bibinfo{journal}{Sci. Rep.} \textbf{\bibinfo{volume}{5}},
  \bibinfo{pages}{7698} (\bibinfo{year}{2015}).

\bibitem[{\citenamefont{Fedorov and Kornyshev}(2008)}]{kornyshev}
\bibinfo{author}{\bibfnamefont{M.}~\bibnamefont{Fedorov}} \bibnamefont{and}
  \bibinfo{author}{\bibfnamefont{A.}~\bibnamefont{Kornyshev}},
  \bibinfo{journal}{J. Phys. Chem. B} \textbf{\bibinfo{volume}{112}},
  \bibinfo{pages}{11868} (\bibinfo{year}{2008}).

\bibitem[{\citenamefont{Capozza et~al.}(2015)\citenamefont{Capozza, Vanossi,
  Benassi, and Tosatti}}]{capozza2015}
\bibinfo{author}{\bibfnamefont{R.}~\bibnamefont{Capozza}},
  \bibinfo{author}{\bibfnamefont{A.}~\bibnamefont{Vanossi}},
  \bibinfo{author}{\bibfnamefont{A.}~\bibnamefont{Benassi}}, \bibnamefont{and}
  \bibinfo{author}{\bibfnamefont{E.}~\bibnamefont{Tosatti}},
  \bibinfo{journal}{J. Chem. Phys.} \textbf{\bibinfo{volume}{142}},
  \bibinfo{pages}{064707} (\bibinfo{year}{2015}).

\bibitem[{\citenamefont{Kim et~al.}(2011)\citenamefont{Kim, Lee, Stoller,
  Dreyer, Bielawski, Ruoff, and Suh}}]{kim11}
\bibinfo{author}{\bibfnamefont{T.~Y.} \bibnamefont{Kim}},
  \bibinfo{author}{\bibfnamefont{H.~W.} \bibnamefont{Lee}},
  \bibinfo{author}{\bibfnamefont{M.}~\bibnamefont{Stoller}},
  \bibinfo{author}{\bibfnamefont{D.~R.} \bibnamefont{Dreyer}},
  \bibinfo{author}{\bibfnamefont{C.~W.} \bibnamefont{Bielawski}},
  \bibinfo{author}{\bibfnamefont{R.~S.} \bibnamefont{Ruoff}}, \bibnamefont{and}
  \bibinfo{author}{\bibfnamefont{K.~S.} \bibnamefont{Suh}},
  \bibinfo{journal}{ACS Nano} \textbf{\bibinfo{volume}{5}},
  \bibinfo{pages}{436} (\bibinfo{year}{2011}).

\bibitem[{\citenamefont{Shim and Kim}(2010)}]{shim10}
\bibinfo{author}{\bibfnamefont{Y.}~\bibnamefont{Shim}} \bibnamefont{and}
  \bibinfo{author}{\bibfnamefont{H.~J.} \bibnamefont{Kim}},
  \bibinfo{journal}{ACS Nano} \textbf{\bibinfo{volume}{4}},
  \bibinfo{pages}{2345} (\bibinfo{year}{2010}).

\bibitem[{\citenamefont{Black et~al.}(2013)\citenamefont{Black, Walters,
  Labuda, Feng, Hillesheim, Dai, Cummings, Kalinin, Proksch, and
  Balke}}]{black13}
\bibinfo{author}{\bibfnamefont{J.}~\bibnamefont{Black}},
  \bibinfo{author}{\bibfnamefont{D.}~\bibnamefont{Walters}},
  \bibinfo{author}{\bibfnamefont{A.}~\bibnamefont{Labuda}},
  \bibinfo{author}{\bibfnamefont{G.}~\bibnamefont{Feng}},
  \bibinfo{author}{\bibfnamefont{P.}~\bibnamefont{Hillesheim}},
  \bibinfo{author}{\bibfnamefont{S.}~\bibnamefont{Dai}},
  \bibinfo{author}{\bibfnamefont{P.}~\bibnamefont{Cummings}},
  \bibinfo{author}{\bibfnamefont{S.}~\bibnamefont{Kalinin}},
  \bibinfo{author}{\bibfnamefont{R.}~\bibnamefont{Proksch}}, \bibnamefont{and}
  \bibinfo{author}{\bibfnamefont{N.}~\bibnamefont{Balke}},
  \bibinfo{journal}{Nano Lett.} \textbf{\bibinfo{volume}{13}},
  \bibinfo{pages}{5954} (\bibinfo{year}{2013}).

\bibitem[{\citenamefont{Dold et~al.}(2015)\citenamefont{Dold, Amann, and
  Kailer}}]{dold2015}
\bibinfo{author}{\bibfnamefont{C.}~\bibnamefont{Dold}},
  \bibinfo{author}{\bibfnamefont{T.}~\bibnamefont{Amann}}, \bibnamefont{and}
  \bibinfo{author}{\bibfnamefont{A.}~\bibnamefont{Kailer}},
  \bibinfo{journal}{Phys. Chem. Chem. Phys.} \textbf{\bibinfo{volume}{17}},
  \bibinfo{pages}{10339} (\bibinfo{year}{2015}).

\bibitem[{\citenamefont{Yang et~al.}(2014)\citenamefont{Yang, Meng, and
  Tian}}]{yang14}
\bibinfo{author}{\bibfnamefont{X.}~\bibnamefont{Yang}},
  \bibinfo{author}{\bibfnamefont{Y.}~\bibnamefont{Meng}}, \bibnamefont{and}
  \bibinfo{author}{\bibfnamefont{Y.}~\bibnamefont{Tian}},
  \bibinfo{journal}{Trib. Lett.} \textbf{\bibinfo{volume}{56}},
  \bibinfo{pages}{161} (\bibinfo{year}{2014}).

\bibitem[{\citenamefont{Smith et~al.}(2013{\natexlab{a}})\citenamefont{Smith,
  Lovelock, Gosvami, Welton, and Perkin}}]{perkin}
\bibinfo{author}{\bibfnamefont{A.}~\bibnamefont{Smith}},
  \bibinfo{author}{\bibfnamefont{K.}~\bibnamefont{Lovelock}},
  \bibinfo{author}{\bibfnamefont{N.}~\bibnamefont{Gosvami}},
  \bibinfo{author}{\bibfnamefont{T.}~\bibnamefont{Welton}}, \bibnamefont{and}
  \bibinfo{author}{\bibfnamefont{S.}~\bibnamefont{Perkin}},
  \bibinfo{journal}{Phys. Chem. Chem. Phys.} \textbf{\bibinfo{volume}{15}},
  \bibinfo{pages}{15317} (\bibinfo{year}{2013}{\natexlab{a}}).

\bibitem[{\citenamefont{Atkin and Warr}(2007)}]{atkin07}
\bibinfo{author}{\bibfnamefont{R.}~\bibnamefont{Atkin}} \bibnamefont{and}
  \bibinfo{author}{\bibfnamefont{G.~G.} \bibnamefont{Warr}},
  \bibinfo{journal}{J. Phys. Chem. C} \textbf{\bibinfo{volume}{111}},
  \bibinfo{pages}{5162} (\bibinfo{year}{2007}).

\bibitem[{\citenamefont{Bou-Malham and Bureau}(2010)}]{bou-malham10}
\bibinfo{author}{\bibfnamefont{I.}~\bibnamefont{Bou-Malham}} \bibnamefont{and}
  \bibinfo{author}{\bibfnamefont{L.}~\bibnamefont{Bureau}},
  \bibinfo{journal}{Soft Matter} \textbf{\bibinfo{volume}{6}},
  \bibinfo{pages}{4062} (\bibinfo{year}{2010}).

\bibitem[{\citenamefont{Ueno et~al.}(2010)\citenamefont{Ueno, Kasuya, Watanabe,
  Mizukami, and Kurihara}}]{ueno10}
\bibinfo{author}{\bibfnamefont{K.}~\bibnamefont{Ueno}},
  \bibinfo{author}{\bibfnamefont{M.}~\bibnamefont{Kasuya}},
  \bibinfo{author}{\bibfnamefont{M.}~\bibnamefont{Watanabe}},
  \bibinfo{author}{\bibfnamefont{M.}~\bibnamefont{Mizukami}}, \bibnamefont{and}
  \bibinfo{author}{\bibfnamefont{K.}~\bibnamefont{Kurihara}},
  \bibinfo{journal}{Phys. Chem. Chem. Phys.} \textbf{\bibinfo{volume}{12}},
  \bibinfo{pages}{4066} (\bibinfo{year}{2010}).

\bibitem[{\citenamefont{Perkin et~al.}(2011)\citenamefont{Perkin, Crowhurst,
  Niedermeyer, Welton, Smith, and Gosvami}}]{perkin11}
\bibinfo{author}{\bibfnamefont{S.}~\bibnamefont{Perkin}},
  \bibinfo{author}{\bibfnamefont{L.}~\bibnamefont{Crowhurst}},
  \bibinfo{author}{\bibfnamefont{H.}~\bibnamefont{Niedermeyer}},
  \bibinfo{author}{\bibfnamefont{T.}~\bibnamefont{Welton}},
  \bibinfo{author}{\bibfnamefont{A.~M.} \bibnamefont{Smith}}, \bibnamefont{and}
  \bibinfo{author}{\bibfnamefont{N.~N.} \bibnamefont{Gosvami}},
  \bibinfo{journal}{Chem. Commun.} \textbf{\bibinfo{volume}{47}},
  \bibinfo{pages}{6572} (\bibinfo{year}{2011}).

\bibitem[{\citenamefont{Perkin}(2012)}]{perkin12}
\bibinfo{author}{\bibfnamefont{S.}~\bibnamefont{Perkin}},
  \bibinfo{journal}{Phys. Chem. Chem. Phys.} \textbf{\bibinfo{volume}{14}},
  \bibinfo{pages}{5052} (\bibinfo{year}{2012}).

\bibitem[{\citenamefont{Li et~al.}(2013{\natexlab{b}})\citenamefont{Li, Endres,
  and Atkin}}]{li13b}
\bibinfo{author}{\bibfnamefont{H.}~\bibnamefont{Li}},
  \bibinfo{author}{\bibfnamefont{F.}~\bibnamefont{Endres}}, \bibnamefont{and}
  \bibinfo{author}{\bibfnamefont{R.}~\bibnamefont{Atkin}},
  \bibinfo{journal}{Phys. Chem. Chem. Phys.} \textbf{\bibinfo{volume}{15}},
  \bibinfo{pages}{14624} (\bibinfo{year}{2013}{\natexlab{b}}).

\bibitem[{\citenamefont{Smith et~al.}(2013{\natexlab{b}})\citenamefont{Smith,
  Lovelock, Gosvami, Licence, Dolan, Welton, and Perkin}}]{smith13}
\bibinfo{author}{\bibfnamefont{A.~M.} \bibnamefont{Smith}},
  \bibinfo{author}{\bibfnamefont{K.~R.~J.} \bibnamefont{Lovelock}},
  \bibinfo{author}{\bibfnamefont{N.~N.} \bibnamefont{Gosvami}},
  \bibinfo{author}{\bibfnamefont{P.}~\bibnamefont{Licence}},
  \bibinfo{author}{\bibfnamefont{A.}~\bibnamefont{Dolan}},
  \bibinfo{author}{\bibfnamefont{T.}~\bibnamefont{Welton}}, \bibnamefont{and}
  \bibinfo{author}{\bibfnamefont{S.}~\bibnamefont{Perkin}},
  \bibinfo{journal}{J. Phys. Chem. Lett.} \textbf{\bibinfo{volume}{4}},
  \bibinfo{pages}{378} (\bibinfo{year}{2013}{\natexlab{b}}).

\bibitem[{\citenamefont{Plimpton}(1995)}]{lammps}
\bibinfo{author}{\bibfnamefont{S.}~\bibnamefont{Plimpton}},
  \bibinfo{journal}{J. Comp. Phys.} \textbf{\bibinfo{volume}{117}},
  \bibinfo{pages}{1} (\bibinfo{year}{1995}).

\bibitem[{\citenamefont{Northern et~al.}(1991)\citenamefont{Northern, Chen,
  Israelachvili, and Zasadzinski}}]{northern91}
\bibinfo{author}{\bibfnamefont{B.~L.~D.} \bibnamefont{Northern}},
  \bibinfo{author}{\bibfnamefont{Y.~L.} \bibnamefont{Chen}},
  \bibinfo{author}{\bibfnamefont{J.~N.} \bibnamefont{Israelachvili}},
  \bibnamefont{and} \bibinfo{author}{\bibfnamefont{J.~A.~N.}
  \bibnamefont{Zasadzinski}}, \emph{\bibinfo{title}{Proceedings of the 49th
  Annual Meeting of the Electron Microscopy Society of America, Atomic force
  microscopy of mica surface after ion replacement, p.628}}
  (\bibinfo{publisher}{San Francisco Press., Inc.}, \bibinfo{address}{San
  Francisco, CA 94101-6800, USA}, \bibinfo{year}{1991}).

\bibitem[{\citenamefont{Segura et~al.}(2013)\citenamefont{Segura, Elbourne,
  Wanless, Warr, Vo\"{i}chovsky, and Atkin}}]{segura13}
\bibinfo{author}{\bibfnamefont{J.~J.} \bibnamefont{Segura}},
  \bibinfo{author}{\bibfnamefont{A.}~\bibnamefont{Elbourne}},
  \bibinfo{author}{\bibfnamefont{E.~J.} \bibnamefont{Wanless}},
  \bibinfo{author}{\bibfnamefont{G.~G.} \bibnamefont{Warr}},
  \bibinfo{author}{\bibfnamefont{K.}~\bibnamefont{Vo\"{i}chovsky}},
  \bibnamefont{and} \bibinfo{author}{\bibfnamefont{R.}~\bibnamefont{Atkin}},
  \bibinfo{journal}{Phys. Chem. Chem. Phys.} \textbf{\bibinfo{volume}{15}},
  \bibinfo{pages}{3320} (\bibinfo{year}{2013}).

\bibitem[{\citenamefont{Elbourne et~al.}(2015)\citenamefont{Elbourne,
  Vo\"{i}chovsky, Warr, and Atkin}}]{elbourne15}
\bibinfo{author}{\bibfnamefont{A.}~\bibnamefont{Elbourne}},
  \bibinfo{author}{\bibfnamefont{K.}~\bibnamefont{Vo\"{i}chovsky}},
  \bibinfo{author}{\bibfnamefont{G.~G.} \bibnamefont{Warr}}, \bibnamefont{and}
  \bibinfo{author}{\bibfnamefont{R.}~\bibnamefont{Atkin}},
  \bibinfo{journal}{Chem. Sci.} \textbf{\bibinfo{volume}{6}},
  \bibinfo{pages}{527} (\bibinfo{year}{2015}).

\bibitem[{\citenamefont{Hayes et~al.}(2011)\citenamefont{Hayes, Borisenko, Tam,
  Howlett, Endres, and Atkin}}]{atkin2}
\bibinfo{author}{\bibfnamefont{R.}~\bibnamefont{Hayes}},
  \bibinfo{author}{\bibfnamefont{N.}~\bibnamefont{Borisenko}},
  \bibinfo{author}{\bibfnamefont{M.}~\bibnamefont{Tam}},
  \bibinfo{author}{\bibfnamefont{P.}~\bibnamefont{Howlett}},
  \bibinfo{author}{\bibfnamefont{F.}~\bibnamefont{Endres}}, \bibnamefont{and}
  \bibinfo{author}{\bibfnamefont{R.}~\bibnamefont{Atkin}}, \bibinfo{journal}{J.
  Phys. Chem. C} \textbf{\bibinfo{volume}{115}}, \bibinfo{pages}{6855}
  (\bibinfo{year}{2011}).

\bibitem[{\citenamefont{Vanossi et~al.}(2013)\citenamefont{Vanossi, Manini,
  Urbakh, Zapperi, and Tosatti}}]{vanossi13}
\bibinfo{author}{\bibfnamefont{A.}~\bibnamefont{Vanossi}},
  \bibinfo{author}{\bibfnamefont{N.}~\bibnamefont{Manini}},
  \bibinfo{author}{\bibfnamefont{M.}~\bibnamefont{Urbakh}},
  \bibinfo{author}{\bibfnamefont{S.}~\bibnamefont{Zapperi}}, \bibnamefont{and}
  \bibinfo{author}{\bibfnamefont{E.}~\bibnamefont{Tosatti}},
  \bibinfo{journal}{Rev. Mod. Phys.} \textbf{\bibinfo{volume}{85}},
  \bibinfo{pages}{529} (\bibinfo{year}{2013}).

\end{thebibliography}

\end{document}